\documentclass[floats,floatfix,showpacs,amssymb,amsmath,prd,twocolumn,superscriptaddress,nofootinbib, aps]{revtex4-2}

\usepackage{graphicx}
\usepackage{dcolumn}
\usepackage{bm}
\usepackage{hyperref}
\usepackage{natbib}

\usepackage{amsmath, amssymb}
\usepackage[usenames, dvipsnames]{xcolor}
\usepackage{multirow}

\newcommand{\araa}{Annu. Rev. Astron. Astrophys.}
\newcommand{\mnras}{Mon. Not. R. Astron. Soc.}
\newcommand{\aj}{Astron. J.}
\newcommand{\jcap}{J. Cosmo. Astropart. Phys.}
\newcommand{\pasp}{Publ. Astron. Soc. Pac.}
\newcommand{\procspie}{Proc. Soc. Photo-Opt. Inst. Eng.}

\newcommand{\e}[1]{\times 10^{#1}}
\newcommand{\sigm}{\tilde{\sigma}/m_{\rm{DM}}}
\newcommand{\fsigm}{\frac{\tilde{\sigma}}{m_{\rm{DM}}}}
\newcommand{\bmid}{\Bigg\vert}

\begin{document}

\preprint{APS/123-QED}

\title{Testing Self-Interacting Dark Matter with Galaxy Warps}

\author{K. Pardo}
\email{kpardo@caltech.edu}
\affiliation{Astrophysical Sciences, Princeton University, Princeton, NJ 08544, USA
}%
\affiliation{
 Astrophysics, University of Oxford, Denys Wilkinson Building, Keble Road, Oxford OX1 3RH, UK
}%
\author{H. Desmond}
\email{harry.desmond@physics.ox.ac.uk}
\affiliation{
 Astrophysics, University of Oxford, Denys Wilkinson Building, Keble Road, Oxford OX1 3RH, UK
}%
\author{P.~G. Ferreira}
\email{pedro.ferreira@physics.ox.ac.uk}
\affiliation{
 Astrophysics, University of Oxford, Denys Wilkinson Building, Keble Road, Oxford OX1 3RH, UK
}%
\date{\today}

\begin{abstract}
Self-interacting dark matter (SIDM) is an able alternative to collisionless dark matter. If dark matter does have self-interactions, we would expect this to cause a separation between the collisionless stars and the dark matter halo of a galaxy as it falls through a dark matter medium. For stars arranged in a disk, this would generate a U-shaped warp. The magnitude of this warping depends on the SIDM cross section, type of self-interaction, relative velocity of galaxy and background, halo structure, and density of the dark matter medium. In this paper, we set constraints on long-range (light mediator) dark matter self-interaction by means of this signal. We begin by measuring U-shaped warps in $3,213$ edge-on disk galaxies within the Sloan Digital Sky Survey. We then forward-model the expected warp from SIDM on a galaxy-by-galaxy basis by combining models of halo structure, density and velocity field reconstructions, and models for the dark matter interactions. We find no evidence for a contribution to the warps from SIDM. Our constraints are highly dependent on the uncertain velocities of our galaxies: for a normalized Rutherford-like cross section we find $\sigm \lesssim 3\e{-13}~\rm{cm}^2/\rm{g}$ at fixed velocity $v = 300~\rm{km/s}$ -- a bound that scales roughly linearly with increasing $v$. In the appendix we translate these bounds into limits on the momentum transfer cross section, finding $\sigma_T(300~\rm{km/s})/m_{\rm{DM}} \lesssim 0.1~\rm{cm}^2/\rm{g}$. We also consider galaxy velocities from the CosmicFlows-3 catalogue. Our limits are stronger than those from dwarf galaxy evaporation, and we show that they scale well with additional data from the next generation of photometric galaxy surveys. Finally, we forecast constraints for contact and intermediate-range interactions that could be achieved with a similar sample of galaxies in cluster environments, where multi-streaming and the fluid approximation are satisfied.
\end{abstract}

\maketitle


\section{Introduction}\label{sec:intro}
Most of the matter in our universe is composed of dark matter (DM). In particular, non-interacting cold dark matter (CDM) can fit the observations of the cosmic microwave background, large scale structure and galactic rotation curves \citep[e.g.,][]{Tegmark2004,Rubin1980}. However, we have yet to detect a DM particle to determine its properties directly \citep[e.g.,][]{Akerib2017, Cui2017, Aprile2018}. There are also possible discrepancies between observations and CDM predictions, stemming mainly from overprediction of power on small scales \citep[see Ref.][for a recent review]{Bullock2017}.

All particles in the Standard Model have non-gravitational interactions, which makes it reasonable to consider such interactions in the dark sector as well. Observationally, self-interacting dark matter (SIDM) could alleviate the possible small-scale CDM issues by redistributing dark matter out of the centers of halos and suppressing small-scale structure formation \citep{SpergelSteinhardt2000}. For velocity-independent interactions, SIDM cross sections per unit DM mass of $\sigma/m \sim 0.1-1\ \rm{cm}^2/\rm{g}$ would be needed to fit the current observations \citep{Rocha2013}. However, there are constraints on SIDM from a wide variety of systems and experiments \citep[for a comprehensive review, see Ref.][]{Tulin2018}. For example, SIDM would lead to the evaporation of halos due to high-momentum-transfer collisions. Thus, the existence of DM halos in dwarf galaxies places constraints on the cross section \citep{Gnedin2001, Kahlhoefer2014}. SIDM would also allow for the spherical relaxation of cluster halos. The observation of elliptical cluster halos places strong limits on the SIDM cross section from cluster ellipticites \citep{Miralda-Escude2002}, although these are disputed \citep{Peter2013}.

SIDM also modifies the distribution of DM in galaxy and galaxy cluster collisions. In the canonical CDM picture, the DM halos do not interact but pass through one another without collision, while the gas shock-heats and decelerates. If DM has self-interactions, then we would expect the DM to experience a drag as well, with a magnitude depending on the interaction cross section. Thus, the centroid of the DM compared to that of the gas could be used to constrain the SIDM cross section. This method has been employed successfully for galaxy cluster collisions, most famously the Bullet Cluster which disfavors interaction cross sections $\sigma/m > 0.7 - 1~\rm{cm}^2/\rm{g}$ \citep{Markevitch2004, Randall2008} (although some simulations find weaker constraints \citep{Robertson2017}).

We can also expect this effect to leave imprints on galaxies falling into clusters. Specifically, we can look for the separation between the centroid of the stars and the DM. The centroid separation technique has been successfully used in simulations \citep{Massey2011}. Unfortunately, a clear detection of this effect in data is challenging due to the weak-lensing accuracy required, as well as other systematics \citep{Harvey2013}. However, recent work has shown that the infall of galaxies into clusters can leave signatures at larger scales \citep{Banerjee2019}. 

Instead, we can search for other signals of this centroid separation. Ref. \cite{Secco2017} recently considered the SIDM dynamics of a disk galaxy falling into a large galaxy cluster. Using numerical simulations, they found that the separation between the DM and stellar centroids should produce a warp in the stellar disk of the galaxy. This would be a U-shaped warp facing in the direction of motion -- a signature difficult to mimic with baryonic effects. The largest warps should occur in galaxies on first infall into galaxy clusters. The dark matter densities are highest in galaxy clusters and the first infall allows for ample time to form the warp before the direction of the drag force changes at periapsis. Out study is the first to use this signal to test SIDM observationally.

Warps are most easily measured in disk galaxies, which are not typically found in galaxy clusters. Although at lower magnitude, warping should occur in \emph{any} galaxy moving in a dark matter medium. Unfortunately, in this case contact interactions would not be expected to leave an appreciable signature. The interaction timescale at low background density is much longer than the time it takes for the background particles to relax into the halo potential. In the contact case, the relevant relative velocity is between coincident DM streams, which are not present in halos that are relaxed with their environments. Only in clusters does such multi-streaming occur. Neither of these restrictions apply to light-mediator models, where interactions are frequent and operate over large distances.

In this paper, we place constraints on the long-range SIDM cross section by measuring the warps of stellar disks. Sec.~\ref{sec:theory} summarizes the theory of dark matter self-interactions and the relevant physical effects that they induce. Sec.~\ref{sec:methods} describes our methods, including our forward-modelling of SIDM warps and measurements of real galaxies. Sec.~\ref{sec:results} gives our results, while Sec.~\ref{sec:discussion} discusses and concludes. An appendix gives our results in terms of the momentum transfer cross section, $\sigma_T$.

\section{Theory}\label{sec:theory}
DM self-interactions will generally induce a drag force on the DM halo of a galaxy traveling through some background over-dense region. The form of the drag force will depend on the type of self-interaction. For a long-range interaction (velocity and angle-dependent), we expect a drag force $\propto \rho_{\rm{bg}}/v^2$ \citep[e.g.,][]{Kahlhoefer2014}. Were the fluid approximation to hold, a contact (velocity-independent) interaction would generate a drag force $\propto \rho_{\rm{bg}}v^2$, where $\rho_{\rm{bg}}$ is the density of the background dark matter and $v$ is the relative velocity between the halo and the background. For intermediate-range interactions (i.e. where the mass of the mediator is close to the mass of the DM particle), we expect a force law between these two cases. 

Other physics will also affect the final force law. For any one collision between particles, there is a probability of the halo particle being ejected. Over time, this leads to an evaporation of the halo, which will damp the drag force. Finally, we expect some velocity dispersion in both the halo and the background. This will cause a distribution of incoming particle velocity directions, further damping the drag force.

In this section, we develop the equations for the expected stellar warp produced by self-interactions between the DM in a galactic halo and a background overdensity. We begin by finding the drag force per particle mass for the three different types of DM self-interactions (long-range, contact, and intermediate-range), along with the modifications due to evaporation and velocity dispersion of the halo. We then describe the warp this produces within the galaxy's stellar disk.

\subsection{Halo deceleration from DM self-interactions}
Consider a halo moving through some background medium with relative velocity $\vec{v}$. We would like to find the force per unit mass on the halo in the direction of $\vec{v}$ from DM self-interactions between particles in the halo and particles in the medium. This drag force law will depend on several factors, such as the angular and velocity dependencies of the self-interaction and the effects of evaporation and velocity dispersion.

\subsubsection{Long-range interactions}\label{sec:longtheory}
Let us first consider interactions arising from a long-range force. For now, focus on a two-particle interaction: one particle from the halo and one from the background overdensity. In the center of mass (COM) frame, the velocity of the halo particle in the direction of the relative velocity will change by:
\begin{equation}\label{eqn-veldiff}
    \delta v_{||} = v\:(\cos \theta -1) \; ,
\end{equation}
where $v \equiv |\vec{v}|$ and $\theta$ is the scattering angle in the COM frame. Note that $\delta v_{||} \leq 0$ always.

The total number of interactions is given by:
\begin{equation}
    dN = \frac{\rho_{\rm{bg}}}{m_{\rm{DM}}} \frac{d\sigma}{d\Omega}v\ dt\ d\Omega \; ,
\end{equation}
where $\rho_{\rm{bg}}$ is the density of the background overdensity and $d\sigma/d\Omega$ is the differential cross section.

The total drag acceleration is given by integrating over all interactions, which can be written as:
\begin{equation}\label{eqn-generaldrag}
    \vec{a}_{\rm{drag}} = \frac{\vec{F}_{\rm{drag}}}{m_{\rm{DM}}} = \frac{\rho_{\rm{bg}}}{m_{\rm{DM}}} v^2\int \frac{d\sigma}{d\Omega} (\cos\theta - 1)\ d\Omega\ \hat{v}
\end{equation}

Long-range interactions describe DM that interacts via a massless mediator, which introduces angle and velocity dependencies to the scattering. We write the differential cross section as:
\begin{equation}
    \frac{d\sigma}{d\Omega} = \frac{\sigma_0\sin\theta}{\left(\frac{v}{c}\right)^4\sin^4\theta} \; ,
\end{equation}
where $\sigma_0 \equiv \alpha_{\rm{DM}}^2/m_{\rm{DM}}^2$, $\alpha_{\rm{DM}}$ is the coupling strength of the interaction and $m_\text{DM}$ is the DM particle mass. This is like the well-known Rutherford scattering formula, except with an extra $\sin \theta$ factor that we use to regularize the total cross section to prevent the momentum transfer cross section from diverging. In the appendix we consider instead a cutoff at the Debye wavelength. Note that the Rutherford cross section is usually written as a function of $\sin(\theta'/2)$, where $\theta'$ is the scattering angle in the frame in which one of the particles is at rest ($\theta = \theta'/2$).

Using Equation~\ref{eqn-generaldrag}, we find that the drag acceleration from long-range interactions is:
\begin{equation}\label{eqn-longrange}
    \vec{a}_{\rm{drag}} =  -\frac{1}{4} \left( \fsigm \right) \rho_{\rm{bg}} \frac{c^4}{v^2}~ \hat{v} \; ,
\end{equation}
where we define an effective cross section 
$\tilde{\sigma} \equiv -16\pi \sigma_0$ 
\citep[see Ref.][for a similar approach]{Kahlhoefer2014}. We use this form so as to match the contact interactions equation (see below).

The drag force from long-range interactions is maximized for \textit{small} relative velocities. Although evaporation will be important for the contact interaction case, the lack of high-momentum-transfer collisions (for suitably small $\tilde{\sigma}/m_\text{DM}$) means that evaporation is negligible for long-range interactions \citep[see, e.g.,][]{Kummer2017}. 

We have assumed that all of the particles in the halo are traveling with velocity $\vec{v}$. More realistically, the particles in the halo will have some velocity dispersion. Ref. \cite{Kummer2017} find that, for a Maxwellian velocity distribution, this leads to a suppression of the drag force, which is well approximated by:
\begin{equation}\label{eqn-veldisp}
\chi_{\rm{p}} = \frac{v^3}{v^3+v_{\rm{disp}}^3}
\end{equation}
where $v_{\rm{disp}}$ is the dispersion velocity of the particles. The background should also have a velocity dispersion, but we ignore this here because it will be small compared to the dispersion of the halo. Our final equation for the long-range drag acceleration is then: 
\begin{equation}
    \vec{a}_{\rm{drag}} =  -\frac{1}{4}\chi_p \left(\fsigm \right) \rho_{\rm{bg}} \frac{c^4}{v^2}\ \hat{v} \; .
\end{equation}

\subsubsection{Contact interactions}
We now turn to the velocity-independent interactions arising from a contact force. This follows the formalism of Sec.~\ref{sec:longtheory}, except with the appropriate (constant) cross section. As Ref. \cite{Secco2017} shows, for an isotropic interaction, this leads to a drag acceleration of the form:
\begin{equation}\label{eqn-secco}
 \vec{a}^{~\rm{contact}}_{\rm{drag}} = -\frac{1}{4} \left(\frac{\tilde{\sigma}}{m_{\rm{DM}}}\right) \rho_{\rm{bg}} v^2\ \hat{v} \; ,
\end{equation}
where $\tilde{\sigma}= \int d\sigma/d\Omega\  d\Omega$ is the total cross section. Since we are assuming an isotropic cross section here, $d\sigma/d\Omega$ is a constant.

However, this does not take into account the effects of evaporation on the halo. Allowing for evaporation, the drag acceleration is modified to \citep{Markevitch2004, Kummer2017}:
\begin{equation}\label{eqn-kummer}
    \vec{a}^{~\rm{contact}}_{\rm{drag}} = - \frac{\chi_{\rm{d}}}{4}\left(\frac{\tilde{\sigma}}{m_{\rm{DM}}}\right) \rho_{\rm{bg}} v^2\ \hat{v} \; ,
\end{equation}
where $\chi_{\rm{d}}$ is the fraction of events that lead to deceleration rather than evaporation. Ref. \cite{Markevitch2004} find this fraction by considering the momentum change per collision and comparing this to the escape velocity of particles in the halo. This gives:
\begin{equation}\label{eqn-decelfraction}
    \chi_{\rm{d}} = 1 - 4\int^1_{\sqrt{x^2/(1+x^2)}}  dy\ y^2 \sqrt{y^2-x^2(1-y^2)} \; ,
\end{equation}
where $x \equiv v_{\rm{esc}}/v$ and $v_{\rm{esc}}$ is the escape velocity for the halo. If we assume a virialized halo, then $v_{\rm{esc}} = 2v_{\rm{disp}}$.

The velocity dispersion correction does not depend on cross section, and thus has the same form as in the long-range case. Our final equation for the contact drag acceleration is then: 
\begin{equation}\label{eqn-contactall}
    \vec{a}^{~\rm{contact}}_{\rm{drag}}= - \frac{1}{4}\chi_{\rm{d}} \chi_{\rm{p}}\left(\frac{\tilde{\sigma}_{\rm{DM}}}{m_{\rm{DM}}}\right) \rho_{\rm{bg}} v^2\ \hat{v} \; .
\end{equation}

As described in Sec.~\ref{sec:intro}, we cannot reliably constrain contact interactions using the warping of field galaxies. We therefore only implement this drag equation in forecasting possible results for galaxies in clusters, in Sec.~\ref{sec:forecast}.

\subsubsection{Intermediate-range interactions}
As our final case, we consider intermediate-range interactions, where the mediator mass can range from massless to infinitely massive (the contact limit). We do this by interpolating the drag acceleration between the two previous cases:

\begin{equation}\label{eq:m}
    \vec{a}^{\ \rm{inter}}_{\rm{drag}} = - \frac{1}{4}\left(\frac{\tilde{\sigma}_{\rm{DM}}}{m_{\rm{DM}}}\right) \rho_{\rm{bg}} v^2 \left(\frac{c}{v}\right)^m \; \hat{v} \:,
\end{equation}
where $0\leq m\leq4$. When $m=0$, this exactly equals the contact case; when $m=4$, this exactly equals the long-range case. We do not assume a particular differential cross section equation, but rather that any such cross section would map onto this form for the drag force. For example, a cross section commonly used for this type of interaction is \citep[e.g.,][]{Kummer2017}:

\begin{equation}\label{eqn-longdiffcs}
    \frac{d\sigma}{d\theta} = \frac{\sigma_0'\sin\theta}{2\left(1+\frac{(v/c)^2}{w^2}\sin^2\theta\right)^2} \; ,
\end{equation}
where $w = m_{\phi}/m_{\rm{DM}}$, $\phi$ is the mediator and $\sigma_0' \equiv (4\pi \alpha_{\rm{DM}}^2m_{\rm{DM}}^2)/m_{\phi}^4$. This reduces to Rutherford scattering for $v \gg w$ and a contact interaction for $v \ll w$. We find that this gives similar results to our interpolating case (see Sec.~\ref{sec:results}).

As with the previous cases, we would like to include the effects of both velocity dispersion and evaporation. The velocity dispersion does not depend on the cross section, so this is trivial to add. However, the evaporation effect requires some more thought. The evaporation fraction calculation requires knowing the differential cross section \citep{Markevitch2004, Kummer2017}. We circumvent this by noting that the evaporation rate should be bracketed by the contact and long-range cases, which means it must be a rapidly decreasing function of the variable $m$ that governs the range of the interaction. We adopt
\begin{equation}\label{eq:evap}
    \chi^{\rm{inter}}_{d} = 1 - (1- \chi_d)\exp[-2m] \; .
\end{equation}
When $m=0$, $\chi^{\rm{inter}}_{d} = \chi_d$; however, when $m=4$, $\chi^{\rm{inter}}_{d} \sim 1$ and there is no evaporation. Unless there is some extra physics that leads to interesting intermediate behavior, the evaporation fraction should smoothly interpolate between the two cases as described approximately by this function.

Thus our final equation for this interaction, including all physics, is
\begin{equation}
    \vec{a}^{\ \rm{inter}}_{\rm{drag}} = - \frac{1}{4}\chi^{\rm{inter}}_{d}\chi_p\left(\frac{\tilde{\sigma}}{m_{\rm{DM}}}\right) \rho_{\rm{bg}} v^2 \left(\frac{c}{v}\right)^m \; .
\end{equation}
Again we only implement this in forecasting, as the warp predictions for field galaxies are unreliable for small $m$.

\subsection{Galaxy warping in SIDM}
We now know the force on the halo from self-interactions. However, we cannot measure the force directly -- we must instead examine its effect on the morphology of the galaxy. In particular, the displacement between the halo and disk induced by dark matter self-interactions sets up a potential gradient across the disk, which warps it into a cup shape. We calculate this warp by considering the difference in acceleration between the disk's center and a general point along the disk, following Ref. \cite{Desmond2018b}.

\begin{figure}[ht]
\includegraphics[width=0.4\textwidth]{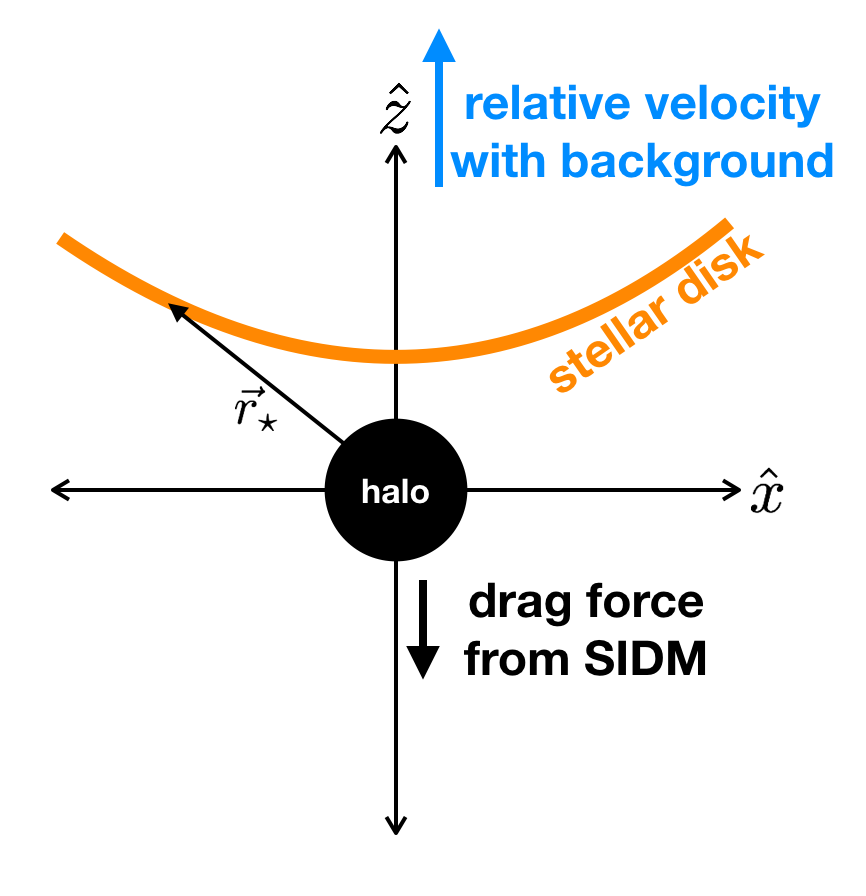}
\caption{Cartoon of how a warp is induced by SIDM. In this picture, the galaxy's stellar disk (orange) and its halo (the center of mass, CM, of the halo is given as the black circle) are falling within an ambient dark matter medium with the relative velocity indicated by the blue arrow. As it falls, the halo experiences a drag force from DM self-interactions, but the stars are collisionless and continue unimpeded. This causes a separation between the centers of the disk and halo, which bends the disk into a U-shaped warp.}
\label{fig-warpdiag}
\end{figure}

Let us define the center of the halo to be at the origin of an $x-z$ plane, where $\hat{z}$ points along the disk normal (see Fig. ~\ref{fig-warpdiag}). The stars are collisionless, but the halo is subject to the drag force derived above. The total acceleration of the halo is:
\begin{equation}
    \vec{a}_{\rm{h}} = \vec{a}_{\rm{bg}} - \vec{a}_{\rm{drag}} \; ,
\end{equation}
where $\vec{a}_{\rm{bg}}$ is the gravitational acceleration due to surrounding matter and $\vec{a}_{\rm{drag}}$ is the drag acceleration due to SIDM. The total acceleration of a point on the stellar disk is:
\begin{equation}
    \vec{a}_\star = \vec{a}_{\rm{bg}} - \frac{GM_{\rm{h}}}{r_\star^2}\hat{r} \; ,
\end{equation}
where $r_\star$ is the equilibrium distance from the point to the center of the halo and $M_h$ is the halo mass enclosed within $r_{\star}$. The second term is the restoring force caused by the offset of the disk from the halo center.

Since we are looking for the equilibrium positions of the stars, we will require that the stars and DM halo move together. This sets $\vec{a}_\star = \vec{a}_{\rm{h}}$, which gives:
\begin{equation}
    \vec{a}_{\rm{drag}} = \frac{GM_{\rm{h}}}{r_\star^2}\hat{z} \; .
\end{equation}

If we assume a spherically-symmetric halo, then the points along the stellar disk will experience different accelerations:
\begin{equation}
    \vec{a}_{\rm{drag}} = \frac{GM_{\rm{h}}}{r_\star^2}\hat{z} \cos\theta = \frac{GM_{\rm{h}}}{r_\star^2}\hat{z} \left(\frac{z}{x}\right) \; .
\end{equation}
We will assume that the warp is slight and thus $x \approx r_{\star}$. This now allows us to write an equation for the $z$ positions of the stars in terms of the drag and the mass of the background halo:
\begin{equation}
    z = a_{\rm{drag}} \frac{|x|^3}{GM_{\rm{h}}}.
\end{equation}
We now assume a power-law density profile for the halo
\begin{equation}
    \rho(r) = \rho_{s} \left( \frac{r_s}{r}\right)^n
\end{equation}
within the extent of the disk, with scale radius $r_s$, $\rho(r_s) \equiv \rho_s$, and a free index $n$ (e.g., $n=1$ for an NFW profile). This gives an enclosed mass $M_h = 4\pi \rho_s/(3-n) r_*^{3-n}$. Substituting this in to our equation for the warp curve above, we find:
\begin{equation}\label{eqn-warpcurve}
    z = a_{\rm{drag}} \frac{3-n}{4\pi \rho_s}|x|^n \; .
\end{equation}

In order to compare to observations, we would like a summary statistic that can quantitatively describe the warp. We will employ the $w_1$ statistic used by Refs.~\cite{Vikram2013} \& \cite{Desmond2018b} -- this is essentially a measure of the average $z$ position across the disk:
\begin{equation}
    w_1 = \frac{2}{L^3} \int_0^L z'x \: dx \; , 
\end{equation}
where $z' = z - \langle z \rangle$, $\langle z \rangle$ is the center of the disk at fixed $x$, and we assume the disk is symmetric about the $z$ axis. Substituting in Equation~\ref{eqn-warpcurve} and integrating, we find
\begin{equation}\label{eqn-w1expected}
    w_1 = \frac{n(3-n)}{(n+1)(n+2)}\frac{a_{\rm{drag}}}{4\pi G\rho_s} \left(\frac{L}{r_s}\right)^n \frac{1}{L} \; .
\end{equation}

\section{Methods}\label{sec:methods}
In this section, we describe the construction of our galaxy sample and explain how we measure the warp curve. Then we describe our model for the estimated warp produced by SIDM.

\subsection{Candidate selection \& warp measurement}
We use the NASA Sloan Atlas (NSA) \citep{Blanton2011} v.1.0.1 catalog\footnote{\url{https://www.sdss.org/dr13/manga/manga-target-selection/nsa/}}, a catalog based mainly on Sloan Digital Sky Survey (SDSS) photometry, to select our galaxies. This catalog contains 641,409 galaxies. General quality cuts (positive mass, radius, flux, and redshift measurements) reduce this number to 640,566. We select only those galaxies that have stellar mass greater than $10^9~ M_{\odot}$ and an axis ratio of $b/a = 0.15$, which leaves us with 22,414 galaxies. This mass cut allows us to use abundance matching to set the dark matter halo masses for our galaxies: the galaxy--halo connection for lower-mass galaxies is considerably more uncertain. The axis ratio cut selects galaxies that are both thin and viewed edge-on. There is some degeneracy between inclination and warp -- an inclined galaxy will always have a smaller warp measurement if we do not properly account for the inclination. Selecting only edge-on galaxies therefore makes the warp curve measurement more robust. We select galaxies within 250\ Mpc, which allows us to use the BORG algorithm to estimate the background density at their positions (see below). Finally, we cut 5 galaxies with defects in their images (cosmic ray streaks across the disk or no galaxy in the $r$-band image at the NSA catalog position or corrupted image file). This leaves a final sample of 3,213 galaxies.

To measure the warp curves we employ the methods of Ref. \cite{Desmond2018b}. We give a short summary of the procedure here. First, we rotate the $r$-band image of a galaxy such that the major axis is aligned with the `$x$-axis'. The warp curve is given by the intensity-weighted $z$ value at each $x$ slice. We then measure the warp using the $w_1$ statistic introduced in Ref. \cite{Vikram2013}:
\begin{equation}
    w_{1,\rm{obs}} = \frac{\int_{-L}^L \frac{x}{L} \frac{z}{L} \frac{dx}{L}}{\int_{-L}^L \frac{x}{L} \frac{dx}{L}} = \frac{1}{L^3}\int_{-L}^L xz\ dx \; ,
\end{equation}
where integration from $-L$ to $L$ allows for asymmetry across the `$z$-axis' (perpendicular to $x$ on the plane of the sky). In practice, we set $L = 3R_{\rm{eff}}$, where $R_{\rm{eff}}$ is the stellar effective radius. We also mask out sky regions for which $|z| > 3 \: b/a \: R_{\rm{eff}} = 0.45 \: R_{\rm{eff}}$.

\subsection{Parameters for estimating the warp}\label{sec:paramwarps}
To calculate the expected warp due to SIDM, we require several pieces of information for each galaxy: the effective radius of the stellar disk ($R_{\rm{eff}}$), the density of the background at the position of the galaxy ($\rho_{\rm{bg}}$), the relative velocity between the galaxy and the background overdensity ($v$), the angle between this relative velocity and the disk normal ($\theta$), the scale radius of the DM halo ($r_s$), the density of the DM halo at the scale radius ($\rho_s$), the power-law index for the DM density profile ($n$), and the velocity dispersion of the halo ($v_{\rm{disp}}$).

We estimate $R_{\rm{eff}}$ by multiplying the measured S\'{e}rsic half-light radius from the NSA catalog, {\ttfamily SERSIC\_TH50}, by the angular diameter distance to the galaxy\footnote{Assuming a flat $\Lambda$CDM cosmology with $h=0.7$, $\Omega_\Lambda = 0.7$, and $\Omega_m = 0.3$.}, with the redshift given by the NSA parameter {\ttfamily ZDIST}. 

We find the halo parameters ($r_s$, $\rho_s$, and $v_{\rm{disp}}$) using halo abundance matching and N-body simulations. Abundance matching (AM) assigns dark matter halos to galaxies by assuming a positive, monotonic relationship between the luminosity or stellar mass of the galaxy and the `proxy', a function of the halo mass and concentration \citep{Kravtsov2004}. Specifically, we use the AM model of Ref.~\cite{Lehmann2017}, which maps the $r$-band absolute magnitude, $M_r$, to a halo proxy given by $v_{\rm{vir}}(v_{\rm{max}}/v_{\rm{vir}})^\alpha$, with a Gaussian scatter $\sigma_\text{AM}$. We take the values $\alpha = 0.6$ and $\sigma_\text{AM} = 0.16~\rm{dex}$, which best reproduce clustering statistics. We use the \textsc{DARKSKY-400} simulation \citep{Skillman2014} post-processed with the \textsc{ROCKSTAR} halo finder \citep{Behroozi2013} for the halo properties. For each matched galaxy--halo pair we calculate $r_s$ and $\rho_s$ from the \textsc{ROCKSTAR} output, assuming an NFW profile \citep{Navarro1997}. Velocity dispersions, $v_{\rm{disp}}$, are calculated by applying the virial theorem to the halos. 

The density of the background, $\rho_{\rm{bg}}$, is estimated from the Bayesian Origin Reconstruction from Galaxies (BORG) algorithm \citep{Jasche2010, Jasche2012, Jasche2013, Jasche2015a, Jasche2015b, Lavaux2016, BorgPM}. This algorithm reconstructs the dark matter density field with a resolution of ${\sim} 2.3~\text{Mpc}/h$ out to ${\sim}250~\text{Mpc}$ by forward-modeling primordial density perturbations with a particle--mesh code and comparing this to the number density field of galaxies in the 2M++ survey \citep{Lavaux2011}. To fill in the smaller-scale power, we also include the mass associated with the 2M++ galaxies themselves, which are linked to halos using the same AM routine as above \citep{Desmond2018maps}.

We use one of two models for galaxy velocities. First, we set $v$ to the same constant (in the range $50 - 10,000~\rm{km/s}$) for all of our galaxies. This is clearly an idealized case, but it gives us a basic idea of the constraining power of our dataset. Second, we use the CosmicFlows-3 (CF3) catalog \citep{Tully2016} of peculiar velocities. We first assign each galaxy a peculiar velocity, $v_{\rm{pec}}$, belonging to the CF3 galaxy closest to it in 3D space. We then assume that the galaxy is falling towards the nearest 2M++ galaxy. The SIDM prediction for the warp we see on the sky is proportional to the relative velocity projected onto the sky. We assign the galaxy velocity in the plane of the sky to be equal to the peculiar velocity.

We must then subtract the velocity of the ambient dark matter medium. We use the public large-scale velocity maps of Ref. \citep{Carrick2015} for this purpose,\footnote{\url{https://cosmicflows.iap.fr/}} evaluated at the positions of our galaxies. These maps are estimated using linear perturbation theory and a reconstruction of the large-scale density in the nearby Universe from the 2M++ catalog. They have resolution 4 Mpc/h,\footnote{This distance is large enough that it corresponds to long (i.e. effectively infinite) range interactions between the halo and its surroundings.} and do not provide uncertainty information. We take each of these background velocities and project them onto the sky. We then subtract this velocity from the total galaxy velocity on the sky. The magnitude of this projected velocity is what we call $v$. We then assign the on-sky angle between this velocity and the disk normal, $\theta$, again assuming that our galaxy is falling towards the nearest 2M++ galaxy. The CF3 peculiar velocities and 2M++ galaxy directions should give us a better idea of the order of magnitude of these relative velocities. However, we also consider fractions $f$ of the relative velocity when we use the CF3 velocities -- from $1- 500\%$. Note that all of our velocities are in the CMB restframe.

We note that the relative velocities we find here are similar to those seen in simulations. In particular, we find that the distribution of fractional velocity differences ($v_{\rm{halo}} - v_{\rm{LS}})/ v_{\rm{halo}}$, with $v_{\rm{halo}}$ the average velocity of DM particles within $R_{\rm{vir}}$ and $v_{\rm{LS}}$ the average velocity of DM particles out to $10~R_{\rm{vir}}$ in the direction of halo velocity) in the Horizon-AGN simulation \citep{Dubois2014} is comparable to that of the galaxies in our model, with $v_{\rm{halo}}$ approximated by $v_\text{CF3}$ and $v_\text{LS}$ from the large-scale velocity reconstruction described above.

\begin{table*}[ht]
\begin{tabular}{|l|c|c|}
\hline
\hline
 Parameter & Source of Uncertainty & Model Used\\
 \hline
 $P(n)$ & Inner DM halo density slope & Uniform prior $n \in [0.5, 1.5]$\\
 $P(\rho_{s}, r_s \mid M_r; \alpha, \sigma_{\rm{AM}})$ & Stochasticity in galaxy--halo connection & 200 mock AM catalogs at fixed $\alpha$ and $\sigma_{\rm{AM}}$\\
 $P(\rho_{\rm{bg}} | \vec{x})$ & Background DM density & 10 draws from BORG posterior\\
$P(v)$ & Galaxy relative velocity & Delta function at set velocity (see Sec.~\ref{sec:paramwarps})\\
 $P(\theta)$ & Unknown relative velocity direction & Delta function at set angle (see Sec.~\ref{sec:paramwarps})\\
 \hline
 \hline
\end{tabular}
\caption{Priors used to find the likelihood of the warp statistic for given $\sigm$ and $m$}
\label{tab:parameters}
\end{table*}

With all of these parameters, we can calculate the predicted $w_1$ statistic for each galaxy using Equation~\ref{eqn-w1expected}, for any given $\sigm$. However, this equation is for a single set of parameter values. We instead want a likelihood function for $w_1$ that takes into account the uncertainties on these parameters. For each parameter, we either set it directly ($v$, $\theta$, $R_{\rm{eff}}$) or we sample over some prior distribution (all the rest). For the halo parameters, we perform the AM step independently 200 times, in each case producing a slightly different galaxy--halo connection due to the stochasticity introduced by $\sigma_\text{AM}$. This generates distributions for $\rho_s$ and $r_s$, separately for each galaxy. We then build our prior for the background density, $\rho_{\rm{bg}}$, by finding the density within BORG at the position of the galaxy, $\vec{x}$, at 10 independent steps of the BORG Markov chain. Finally, we use a uniform prior for $n$ from 0.5 to 1.5 independently for each galaxy. This range is chosen to include the NFW value ($n=1$) as well as profiles that are slightly shallower or steeper.

We then perform Monte Carlo sampling for each galaxy independently to determine the $w_1$ likelihood function. Since $w_1 \propto \sigm$, we can generate the likelihood function at $\sigm = 1\ \rm{cm}^2/\rm{g}$ and then simply scale it up or down when sampling $\sigm$:
\begin{widetext}
\begin{align}
     \mathcal{L}\left( w_1 \bmid \fsigm = 1~\rm{cm}^2/\rm{g}, m\right) &= \int \mathcal{L}\left( w_1 \bmid \fsigm = 1~\rm{cm}^2/\rm{g}, m, \rho_s, r_s, n, \rho_\text{bg}, v, \theta \right) \mathcal{L}\left( \rho_s, r_s \mid M_r; \alpha, \sigma_{\rm{AM}}\right)\\ \nonumber
     &\times  \mathcal{L}(\rho_{\rm{bg}}, v \mid \vec{x}) \; P(\theta) P(n)~d\rho_s~dr_s~d\rho_{\rm{bg}}~dv~d\theta~dn \; ,
\end{align}
\end{widetext}
where the probability distributions for each of these priors is given in Table~\ref{tab:parameters}. We test for convergence of the likelihood function for each galaxy by requiring that the mean, variance, and skew of $\mathcal{L}(w_1 | \tilde{\sigma}/m_\text{DM} = 1~\rm{cm}^2/\rm{g}, m)$ does not change by more than 1\% in the last 10\% of the samples, which we find requires at least 100,000 Monte Carlo draws from the prior distributions. Note that by building these distributions directly into the likelihood we are effectively sampling from the priors in these quantities rather than the posteriors, which would be computationally too expensive. While this likelihood is written for general $m$, we remind the reader that the inference is most reliable for larger $m$, corresponding to a longer-range interaction.

\subsection{Parameter inference}

We now have a measured warp statistic for each galaxy, $w_{1,\rm{obs}}$, and the likelihood of a given warp statistic under an SIDM model with $\sigm = 1\ \rm{cm}^2/\rm{g}$. This enables us to derive constraints on $\sigm$ and $m$ using Bayes' theorem and a Markov Chain Monte Carlo (MCMC) algorithm. Note that Equation~\ref{eqn-w1expected} is linear in $\sigm$ and in the other factors (besides $m$) that affect the physics of the interactions. Thus, for our parameter estimation of $\sigm$, we simply sample from $\mathcal{L}(w_1 | \sigm = 1~\rm{cm}^2/\rm{g}, m)$ and then scale by the particular $\sigm$ value the Markov chain is sampling. We then compare this to the measured $w_{1,\rm{obs}}$ value for each galaxy, as described below. In the long-range and contact cases, we fix $m$ at the appropriate values and do not sample over it.

For the most part, the measured warp values are many orders of magnitude larger than the estimated warp parameters, given a reasonable cross section. In other words, noise dominates the warp signal. Given that we have no reasonable model for how other processes may produce U-shaped warps, we assume that the noise is normally distributed and marginalise over its variance, $\sigma_{w_1}^2$. This modifies the $w_1$ likelihood to:
\begin{align}
    \mathcal{L}\Bigg(w_{1,\rm{obs}} &\bmid \fsigm, m, \sigma_{w_1}\Bigg) = \int \frac{dw_1}{\sqrt{2\pi \sigma_{w_1}^2}} \; \\
    &\times \mathcal{L}\left(w_1\bmid\fsigm, m\right)\exp\left[\frac{-(w_{1,\rm{obs}} - w_1)^2}{2 \sigma_{w_1}^2}\right] \; , \nonumber
\end{align}
In practice, we evaluate this integral by discretizing $w_1$ into 50 bins between its minimum and maximum values, separately for each galaxy.

We sample this likelihood using the \emph{emcee} affine-invariant Markov sampler \citep{Foreman-Mackey2013}. We set the flat prior $\sigm \in (0, 10^4)$ and check that varying this prior does not significantly change the results. For the intermediate-range results, we sample $\log_{10} (\sigm/[\rm{cm}^2\ \rm{g}^{-1}]) \in (-20, 2)$ with a uniform prior in log. The power-law index, $m$, has a flat linear prior over $m\in [0,4]$. Finally, we sample in $\log_{10}{\sigma_{w_1}}$, with no restrictions on its range.

For the long-range and contact interactions, we use 10 walkers and take 20,000 samples, after burn-in. This gives a Gelman-Rubin convergence parameter $R < 0.01$. For the intermediate-range case we require $\gtrsim 25,000$ samples after burn-in to give the same level of convergence.

\section{Results}\label{sec:results}

\subsection{Long-range interactions}
\label{sec:main_results}

Our main results are:
\begin{enumerate}
    \item There is no preference for SIDM ($\tilde{\sigma}/m_\text{DM} > 0$) over the null hypothesis that warps are generated purely by astrophysical or measurement noise. This indicates no net positive correlation between the direction of the warps and the galaxies' velocities relative to the background on the plane of the sky, or between the warp magnitude and the expectation of Eq.~\ref{eqn-w1expected}.\footnote{This is predictable from the results of Ref. \cite{Desmond2018b}, who show that there is a negative correlation between the warp direction and the orientation of the fifth-force field in thin-shell-screened modified gravity theories, which is largely aligned with galaxies' velocities. This is the expectation in such theories.}
    \item For long-range interactions, we place a limit of $\sigm (v = 300~\rm{km/s}) < 3\e{-13}~\rm{cm}^2/\rm{g}$, scaling as ${\sim}v^{1.0}$ assuming a constant velocity $v<1000~\rm{km/s}$. This scaling goes as ${\sim}v^{-0.028}$ for the velocities set using fractions of the CF3 velocities. Including the possibility of variation in the galaxy velocities, we find a range of $68\%$ upper bounds on the cross section from $\sigm \lesssim 2\times 10^{-13} - 10^{-10}~\rm{cm}^2/\rm{g}$, with assumed median galaxy velocity from $v\sim 50~\rm{km/s} -  10^{4}~\rm{km/s}$.
    \item For intermediate-range interactions we find a slight preference for smaller $m$, with a corresponding constraint on the cross section. 
\end{enumerate}

\begin{figure}
    \centering
    \includegraphics[width=0.5\textwidth]{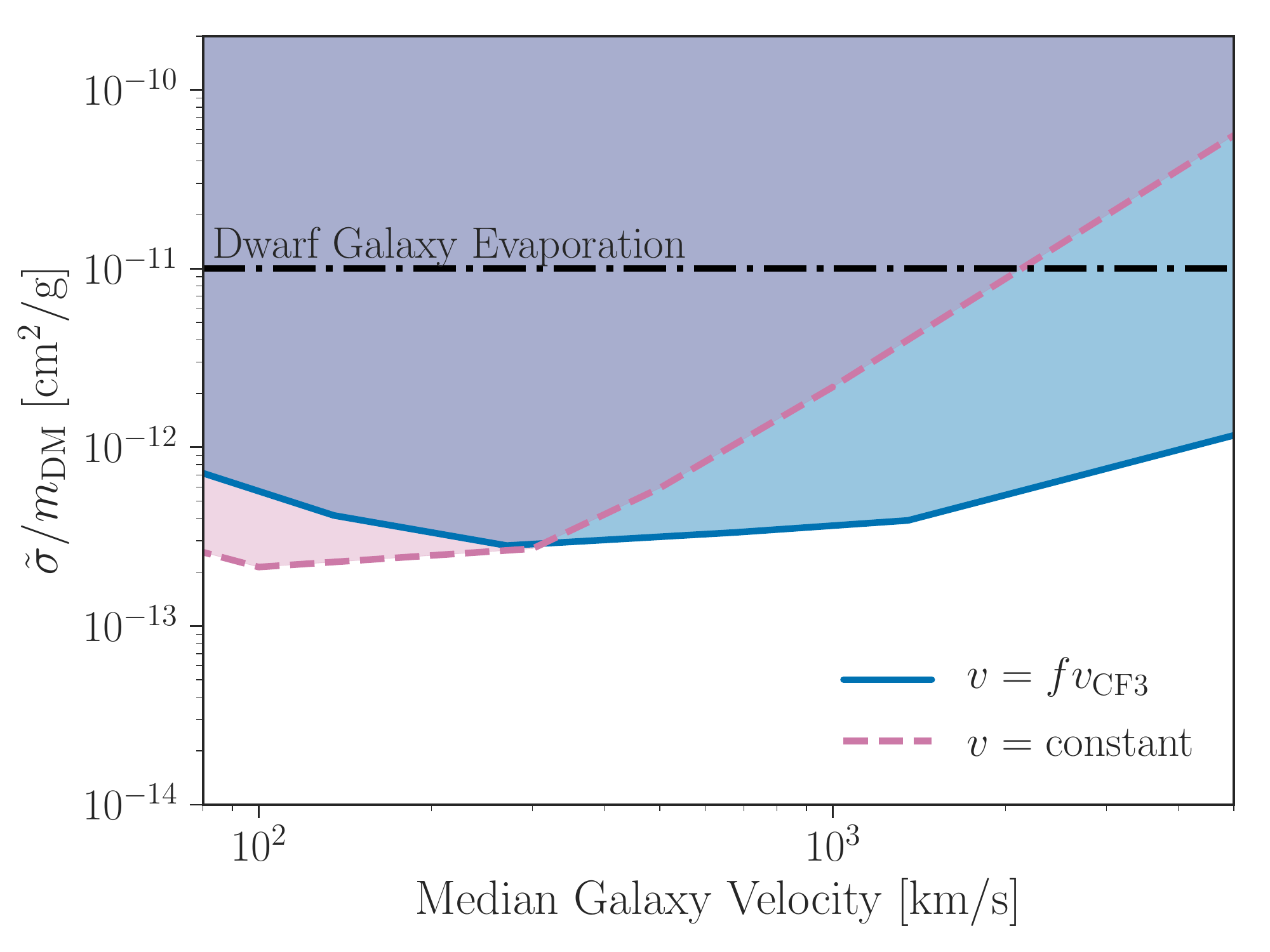}
    \caption{68\% upper limits on the SIDM cross section assuming a long-range interaction versus the median assumed velocity. We show limits assuming either that all galaxies have the same relative velocities (dashed pink), or that they have velocities proportional to their CF3 velocities (solid blue). The black dotted-dashed line gives an upper limit from dwarf galaxy evaporation rates \citep{Kahlhoefer2014}.}
    \label{fig:long_results}
\end{figure}

In the rest of this section, we use different assumptions about the relative velocities of our galaxies to give more detailed results. These results are summarized in Fig. \ref{fig:long_results}, and Table~\ref{tab:results}. Note that in all cases we marginalize over the variance of the noise term, $\sigma_{w_1}$. We find that $\sigma_{w_1}$ is not degenerate with any other model parameter and its posterior is invariant for all of the models we consider. It is peaked at the measured variance of $w_{1,\rm{obs}}$, indicating that it picks up the overall magnitude of the measured warps. The constraints on SIDM parameters instead depend on the correlation of $\hat{w}_1$ with environment and galaxy/halo properties.

Our long-range limits are given in Fig.~\ref{fig:long_results}. As described in Sec.~\ref{sec:paramwarps}, we use either of two assumptions for the velocities: 1) we set all velocities to the same value; or 2) we set the velocities to some fraction of the measured velocities from the CF3 data. The limit on the cross section differs by at most a factor $\sim2$ between these models. The results in Fig.~\ref{fig:long_results} include the effects of velocity dispersion but not evaporation (Sec.~\ref{sec:theory}). We find our limits to be considerably stronger than those from dwarf galaxy evaporation \citep{Kahlhoefer2014} for all but very highest galaxy velocities.\footnote{Note that the evaporation considered in Ref. \citep{Kahlhoefer2014} is different to what we discuss in Sec.~\ref{sec:theory}: they consider evaporation due to long-range interactions over a very long timescale (``cumulative'' evaporation). In addition, evaporation is more pronounced for dwarf galaxies than the larger-mass galaxies we consider.}

\begin{table*}[ht]
\begin{tabular}{|l|l|l||c|c|}
\hline
\hline
 Assumed Velocity & Evaporation? & Dispersion? & 68\% Upper Limit & 95\% Upper Limit\\ & & & $\rm{cm}^2/\rm{g}$ & $\rm{cm}^2/\rm{g}$ \\
 \hline
  \multirow{2}{*}{$v = 300 \: \rm{km/s}$} & N/A & - & $2.0\e{-13}$ & $4.4\e{-13}$\\
  {} & N/A & \checkmark & $2.7\e{-13}$ & $6.1\e{-13}$\\
 \cline{1-5}
 \multirow{2}{*}{$v = v_{\rm{CF3}}$} & N/A & - & $4.7\e{-14} $& $1.2\e{-13}$\\
 {} & N/A & \checkmark & $3.9\e{-13}$ & $9.4\e{-13}$\\
 \hline
 \hline
\end{tabular}
\caption{Limits on the self-interaction cross section $\tilde{\sigma}/m_\text{DM}$ for long-range interactions, for different assumed galaxy velocities}
\label{tab:results}
\end{table*}

\subsection{Forecasted constraints for contact \& intermediate-range interactions}
\label{sec:forecast}

As discussed in Sec.~\ref{sec:intro} (and described further in Sec.~\ref{sec:discussion}), we cannot place reliable limits on contact interactions with our current sample of galaxies. Because of the low background densities, there are few interactions between the halo and background DM particles before they are able to relax to the same average velocity. In other words, there is no difference in velocity between the halo and the background particles in the contact interaction case, eliminating the expected drag. In addition, we find that our intermediate results slightly prefer the contact interactions case, so we also cannot give reliable limits in this case with our current sample.

To give an idea of the contact and intermediate range constraints that \emph{could} be achieved with galaxies in clusters, we repeat our analysis using the corresponding drag forces. We use the same galaxy parameters as in the long-range case, but change the cross section and its prior. This is a conservative forecast in that the background densities (and relative velocities) of our galaxies are significantly lower than in clusters, leading to underestimation of the drag force. The cross section constraints should scale as $\rho_{\rm{bg}}^{-1}$. The average background density for our galaxies is $\rho_{\rm{bg}} = 330~M_{\odot}/\rm{kpc}^3 \sim 2.4\:\rho_{\rm{crit}}$. For galaxies in clusters we would expect $\rho_{\rm{bg}} > 200~\rho_{\rm{crit}}$, which would strengthen constraints by a factor $\gtrsim100$. However, these limits may underpredict the amount of evaporation. Note that the degree of evaporation depends only on the ratio of the relative velocity to the escape velocity of particles within the halo. Thus evaporation effects would only become overwhelming if the relative velocities were, on average, many times larger than the escape velocities. Given that the escape velocity in a typical disk galaxy is $>500~\rm{km/s}$ and the dispersion of galaxies in clusters is $\sim 1000~\rm{km/s}$, we do not expect this to be the case. Our forecasts should therefore provide conservative upper limits for our sample size, although we caution that it will be observationally challenging to find this many thin disks in cluster environments.

\begin{figure}
    \centering
    \includegraphics[width=0.5\textwidth]{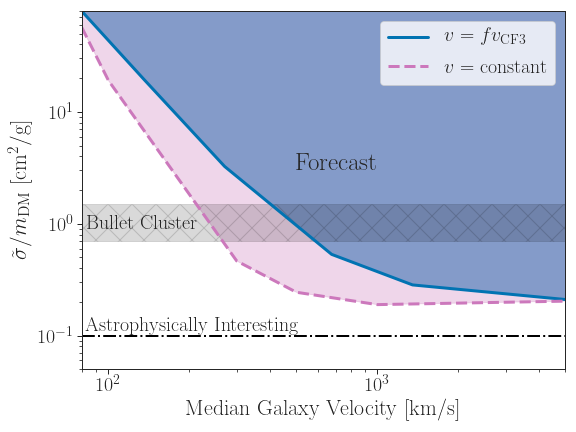}
    \caption{Forecasted 68\% upper limits on the SIDM cross section assuming a contact interaction versus the median assumed velocity, assuming a sample similar to ours but in environments where multi-streaming and the fluid approximation obtain. We show limits assuming either that all galaxies have the same relative velocities (pink), or that they have velocities proportional to their CF3 velocities (blue). The grey hatched region gives the range of constraints on the cross section from the Bullet Cluster \citep{Markevitch2004, Randall2008, Kahlhoefer2014}. The dotted-dashed line gives the minimum SIDM cross section needed to provide astrophysically interesting effects (i.e. suppression of small scale structure and DM halo cores) \citep{Rocha2013}.}
    \label{fig:contact_results}
\end{figure}

\begin{figure}
    \centering
    \includegraphics[width=0.5\textwidth]{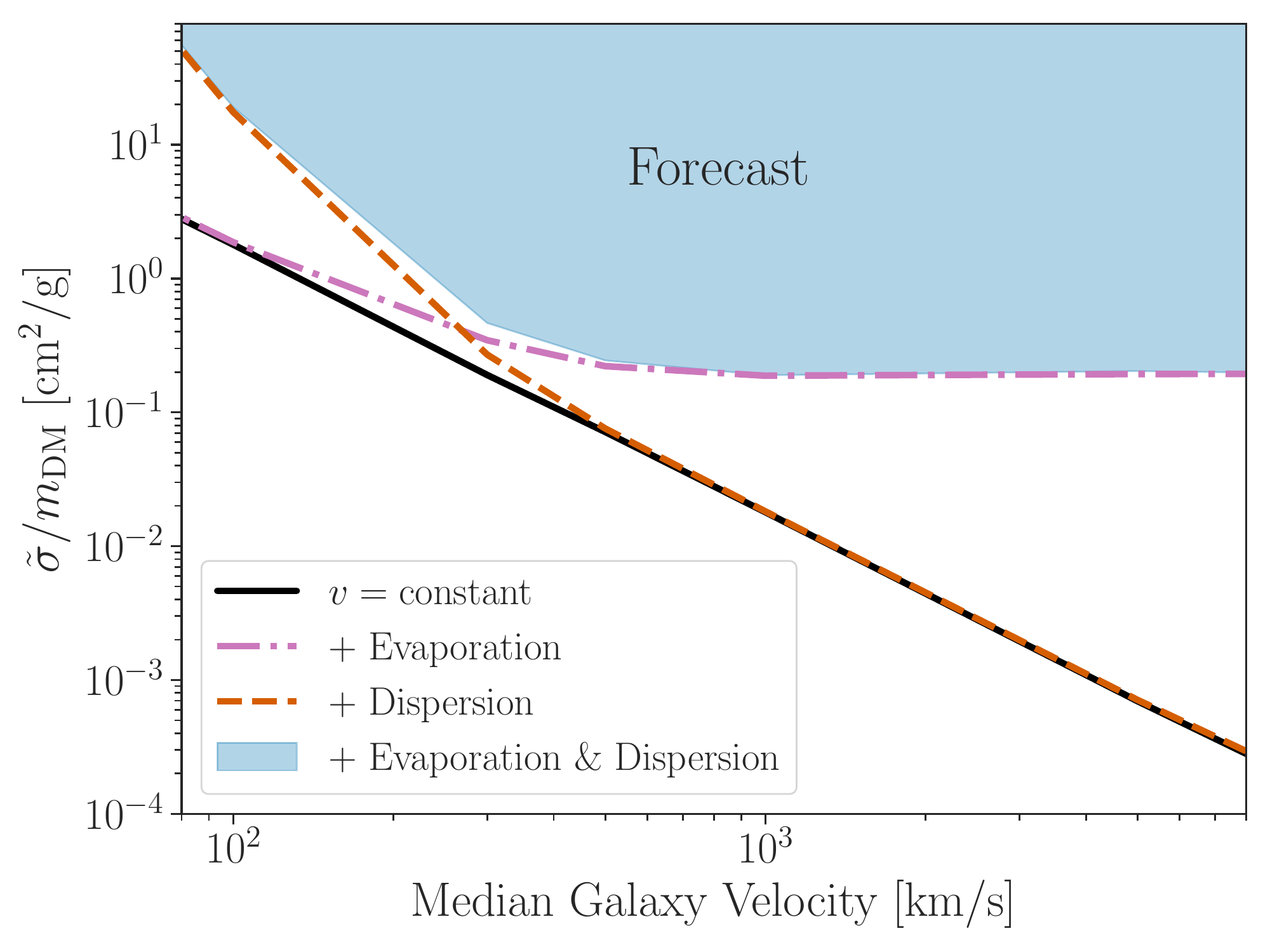}
    \caption{Forecasted 68\% upper limits on the SIDM cross section assuming a contact interaction versus the median assumed velocity, as Fig.~\ref{fig:contact_results}. Here we show the effects of velocity dispersion and evaporation on the results. The black line shows the limits if we do not consider either of these physical effects. The pink, dotted-dashed line includes evaporation and the orange, dashed line includes velocity dispersion. The blue region shows the same limits as Fig.~\ref{fig:contact_results}, which includes both effects.}
    \label{fig:evap_veldisp_effects}
\end{figure}

\begin{figure}
    \centering
    \includegraphics[width=0.5\textwidth]{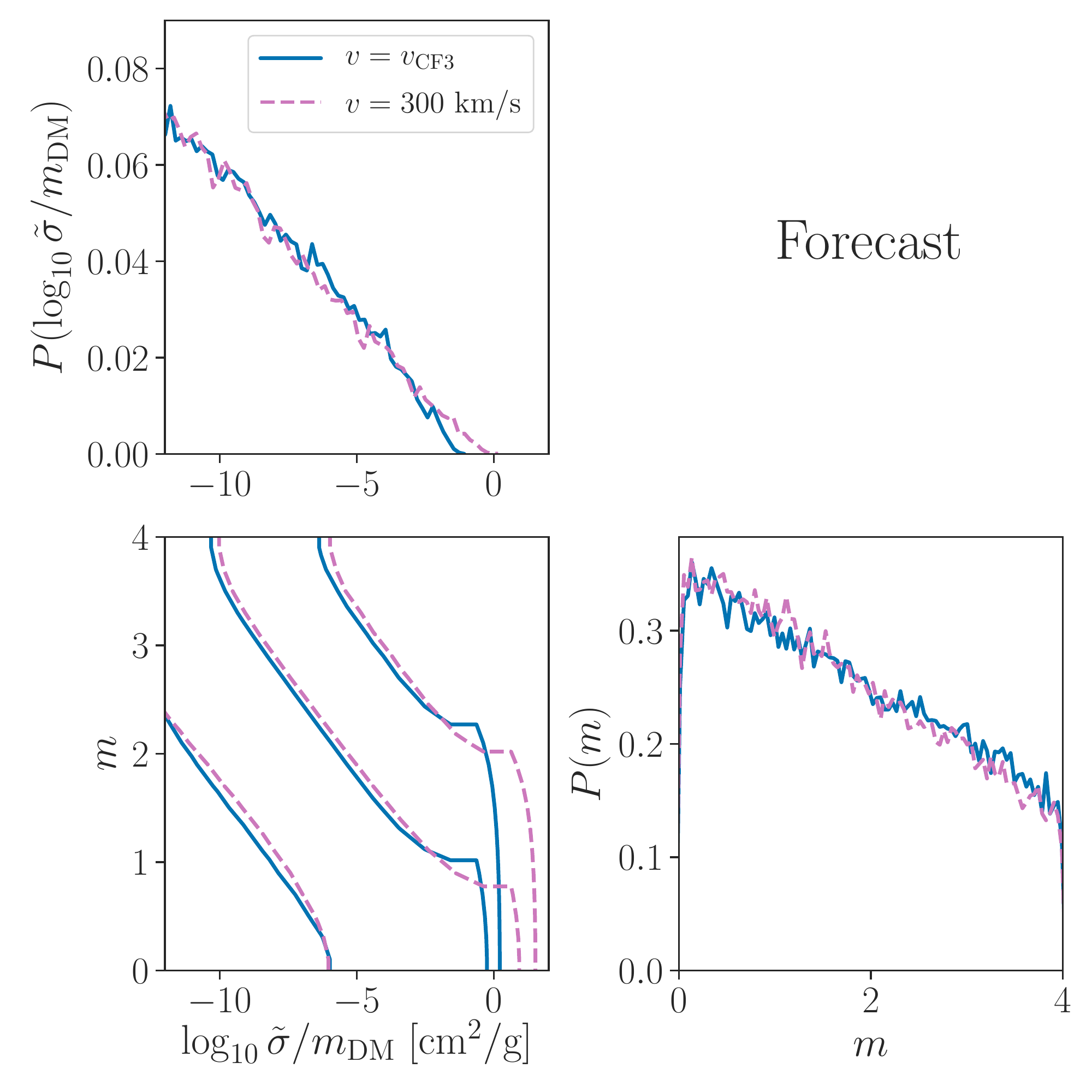}
    \caption{Forecasted corner plot for an intermediate-range DM self-interaction, assuming a sample similar to ours but in environments where multi-streaming and the fluid approximation obtain. We show our limits assuming all galaxies have $v=300~\rm{km/s}$ (pink) and assuming they have velocities set by their CF3 velocities (blue). $m$ determines the dependence of $a_\text{drag}$ on the relative velocity of the halo and background (Eq.~\ref{eq:m}). Note that because we use a Jeffrey's prior here for $\sigm$ and the posterior peaks at $\sigm = 0~\rm{cm}^2/\rm{g}$, the confidence levels depend sensitively on the arbitrary lower limit of the prior and should not be used: the contour lines in the off-diagonal panel are meant merely to show the degeneracy direction.}
    \label{fig:interp_corner}
\end{figure}

We give our main results in Fig.~\ref{fig:contact_results}. As with the long-range case, we report our limits as a function of the assumed velocity and show both the constant and CF3 velocity models. These limits for average velocities greater than $\sim 500~\rm{km/s}$ are tighter than the Bullet Cluster constraints \citep{Markevitch2004, Randall2008, Kahlhoefer2014}. In Fig.~\ref{fig:evap_veldisp_effects}, we show how the evaporation and velocity dispersion effects change our limits. Adding both of these effects (as is done in Figs.~\ref{fig:interp_corner} and~\ref{fig:contact_results}) weakens the limits by about one order of magnitude, regardless of the velocity scale.

Finally, we consider the intermediate case in Fig.~\ref{fig:interp_corner}. This shows the posterior distributions for $\sigm$ and $m$, the power-law index for the velocity dependence of the interaction. Contact interactions (low $m$ values) are slightly preferred, although this may be solely because they allow a larger volume of the $\tilde{\sigma}/m_\text{DM}$ prior. In addition, the very low $m$ values are unreliable -- they suffer the same issues as the contact interactions case (see Sec.~\ref{sec:discussion}). Note that we use a logarithmic prior on the cross section in this case due to the enormous width of the posterior as $m$ varies. However, since the posterior peaks at $\sigm = 0~\rm{cm}^2/\rm{g}$, confidence limits depend on the arbitrary lower limit of the prior and are therefore not reliable. The shapes of the posteriors and their dependence on velocity are nevertheless robust.

\section{Discussion}\label{sec:discussion}

Our results in the previous section show that we can place new constraints on SIDM cross sections by measuring the warps of stellar disks. In this section, we discuss possible systematics and our attempts to mitigate them. We also discuss the prospects for improving the constraints with next-generation surveys.

As shown in Figs.~\ref{fig:long_results} and \ref{fig:contact_results}, the bounds on the cross section are dictated by the magnitudes of the galaxies' relative velocities. We have provided a range of constraints based on different reasonable assumptions, but more robust limits require more precise velocity measurements. The CF3 velocities have very large errors, in excess of 100\% at times. In addition, the CF3 catalog does not include most of the galaxies in our sample, forcing us to assign velocities by means of a nearest neighbour algorithm. Most of our sample is within $\sim 10~\rm{Mpc}$ of a CF3 galaxy. We find that, within the CF3 catalog, the velocities are well-correlated on these scales. We therefore expect this to be an adequate estimator of the true velocity, but caution that it must introduce some uncertainty. Note also that we do not include uncertainties on the peculiar velocities in our likelihood function. 

Another possible systematic is the effect of baryonic physics on galaxy morphology. Most warps caused by tidal or baryonic effects are S-shaped \citep{Binney1992}, and are therefore effectively filtered out by our choice of warp statistic. Any non-SIDM contribution to $w_1$ is captured to leading order by our noise model (marginalization over $\sigma_{w_1}$), but only under the assumption that this contribution is Gaussian and independent of environment and galaxy/halo properties. Baryonic and tidal effects are likely to break this assumption to some degree. In addition, gas in the galaxy will experience hydrodynamical drag from interaction with gas in the intergalactic medium (IGM), which will lead to a U-shaped warp in the same direction as SIDM. Thus, including this IGM contribution would tighten our limits, making our results again conservative. The location of the gas as well as the dependence of the measured warp on gas mass would help break the degeneracy between these two types of physics in the context of future, more precise constraints.

We also neglect the effects of tidal interactions, which could contribute to anisotropy in halo and galaxy profiles. These would be largest within clusters while our galaxies are mainly in the field, so we do not expect it to significantly bias our results. However, future constraints on the contact interaction would require a sample of galaxies within clusters, where tidal interactions may need to be considered more carefully.

On the theory side, we use the fluid approximation to derive the SIDM prediction for the warp. As we have mentioned, given the low background densities of our field galaxies, the fluid approximation is not valid for contact interactions. The average background densities near our galaxies is $\rho_{\rm{bg}} = 330~M_{\odot}/\rm{kpc}^3 \sim 2.4 \rho_{\rm{crit}}$, where $\rho_{\rm{crit}}$ is the critical density today. For $v=300\:\rm{km}/{s}$ and $\sigm = 1~\rm{cm}^2/\rm{g}$ this gives an interaction time larger than $1/H_0$. In this time, most of the background particles would easily have time to relax into the potential of the halo, giving them the same velocity as the halo particles. There would then be no relative velocity between the background and halo particles, and no drag effect. If our galaxies were located in clusters, the background density would be high enough to ensure interactions could occur before relaxation. In other words, a warping effect from short-range interactions requires multi-streaming. The long-range results also depend on the fluid approximation, but there are many more interactions because of the nature of the force: the interaction times in this case are closer $\sim 50~\rm{Myr}$, which is less than the typical dynamical times for these galaxies.

The finite time required for thermalization may have an impact on the warp shape. If the central parts of the halo thermalize faster than the outer parts due to a greater interaction rate the inner halo will experience a stronger drag force, leading to an asymmetry with respect to the dark matter further out. The failure of the halo to move in one piece may bias our warp model. This effect is likely to depend on the total interaction rate (i.e., the cross section, background density, and velocities), thus numerical simulations of galaxies in similar environments to ours would be required to assess its magnitude.


Finally, we neglected the self-gravity of the disk in our calculations. Ref.~\cite{Desmond2018b} found this to be a small effect. However, as imaging and analysis techniques improve this may become a relevant systematic.

In the upcoming era of large and deep photometric surveys (e.g., LSST\footnote{\url{https://www.lsst.org/}} \citep{LSST2009}, WFIRST\footnote{\url{https://wfirst.ipac.caltech.edu/}} \citep{Spergel2013} and Euclid\footnote{\url{https://www.euclid-ec.org/}} \citep{Laureijs:2012}) we can expect to have a much larger sample of edge-on galaxies to test. Assuming that we can continue to measure the properties of the DM background (density and velocity) in these survey volumes, as well as the galaxies' peculiar velocities, we can expect these samples to yield considerably tighter constraints. To quantify this, we repeat our analysis for the long-range interaction case (for $v=300~\rm{km/s}$ and without the velocity dispersion effect) using random subsets of size $N$ of our galaxy sample. This produces a range of results depending on the subset of galaxies chosen. For each subset size, we run 1000 separate MCMC chains and record the 68\% upper limits on $\sigm$ for each chain. We find that the upper 16\% of these limits is well fit by $\tilde{\sigma}/m_{\rm{DM}}|_{1\sigma} \propto N^{-0.9} \;$. In other words, in the worst case scenario that all of the future galaxies have the same constraining power as our least-constraining few hundred galaxies, with 10,000 galaxies the limits would be tighter than those of Sec.~\ref{sec:forecast} by a factor of $\sim2.8$. The median limits show that with this same number we can more likely expect at least an order of magnitude better constraints. This is even without accounting for any improvements in the velocity determination and other modeling. We can further improve these constraints by finding more thin, edge-on galaxies in high density environments, which would be expected to have the largest warp signature and thus the greatest constraining power. We would also want to choose galaxies at low redshift and with high stellar mass, reducing uncertainties in both measuring the warps and assigning halo properties to the galaxies.

In summary, we calculate the expected stellar disk warp due to DM self-interactions for a variety of interaction types and additional physical processes. We then compare these to the measured warps of edge-on disk galaxies in the SDSS to place constraints on long-range interactions that are stronger than those from dwarf galaxy evaporation. These results are conservative given our treatment of the interstellar medium and velocity uncertainties, although there remain modeling challenges (e.g. the use of the fluid approximation and the precise values of galaxies' peculiar velocities). We also show that a similar sample of galaxies in cluster environments would place highly competitive constraints on contact and intermediate-range interactions. Given the power of this probe, we believe this to be a fruitful avenue for future work. With more galaxies, better photometry and more accurate velocities, we can hope to use galaxy structure either to detect SIDM, or to rule it out as an astrophysically interesting possibility.

\acknowledgements

We are very grateful to Michael Blanton for sharing the NSA v1.0.1 images necessary for this work. We thank James Binney, Hume Feldman, Michael Hudson, Bhuvnesh Jain, Manoj Kaplinghat, Andrew Robertson, Lucas Secco, David Spergel, and Richard Watkins for useful discussions, and an anonymous referee for comments which greatly improved the manuscript. KP received support from the Balzan foundation via New College, Oxford, and from the National Science Foundation Graduate Research Fellowship Program  under grant DGE-1656466. HD is supported by St John's College, Oxford, and acknowledges financial support from ERC Grant No 693024 and the Beecroft Trust. PGF is supported by the ERC, STFC, and the Beecroft Trust.

\bibliography{ref}

\begin{thebibliography}{47}%
\makeatletter
\providecommand \@ifxundefined [1]{%
 \@ifx{#1\undefined}
}%
\providecommand \@ifnum [1]{%
 \ifnum #1\expandafter \@firstoftwo
 \else \expandafter \@secondoftwo
 \fi
}%
\providecommand \@ifx [1]{%
 \ifx #1\expandafter \@firstoftwo
 \else \expandafter \@secondoftwo
 \fi
}%
\providecommand \natexlab [1]{#1}%
\providecommand \enquote  [1]{``#1''}%
\providecommand \bibnamefont  [1]{#1}%
\providecommand \bibfnamefont [1]{#1}%
\providecommand \citenamefont [1]{#1}%
\providecommand \href@noop [0]{\@secondoftwo}%
\providecommand \href [0]{\begingroup \@sanitize@url \@href}%
\providecommand \@href[1]{\@@startlink{#1}\@@href}%
\providecommand \@@href[1]{\endgroup#1\@@endlink}%
\providecommand \@sanitize@url [0]{\catcode `\\12\catcode `\$12\catcode
  `\&12\catcode `\#12\catcode `\^12\catcode `\_12\catcode `\%12\relax}%
\providecommand \@@startlink[1]{}%
\providecommand \@@endlink[0]{}%
\providecommand \url  [0]{\begingroup\@sanitize@url \@url }%
\providecommand \@url [1]{\endgroup\@href {#1}{\urlprefix }}%
\providecommand \urlprefix  [0]{URL }%
\providecommand \Eprint [0]{\href }%
\providecommand \doibase [0]{https://doi.org/}%
\providecommand \selectlanguage [0]{\@gobble}%
\providecommand \bibinfo  [0]{\@secondoftwo}%
\providecommand \bibfield  [0]{\@secondoftwo}%
\providecommand \translation [1]{[#1]}%
\providecommand \BibitemOpen [0]{}%
\providecommand \bibitemStop [0]{}%
\providecommand \bibitemNoStop [0]{.\EOS\space}%
\providecommand \EOS [0]{\spacefactor3000\relax}%
\providecommand \BibitemShut  [1]{\csname bibitem#1\endcsname}%
\let\auto@bib@innerbib\@empty
\bibitem [{\citenamefont {{Tegmark}}\ \emph {et~al.}(2004)\citenamefont
  {{Tegmark}}, \citenamefont {{Strauss}}, \citenamefont {{Blanton}},
  \citenamefont {{Abazajian}}, \citenamefont {{Dodelson}}, \citenamefont
  {{Sandvik}}, \citenamefont {{Wang}}, \citenamefont {{Weinberg}},
  \citenamefont {{Zehavi}},\ and\ \citenamefont {{Bahcall}}}]{Tegmark2004}%
  \BibitemOpen
  \bibfield  {author} {\bibinfo {author} {\bibfnamefont {M.}~\bibnamefont
  {{Tegmark}}}, \bibinfo {author} {\bibfnamefont {M.~A.}\ \bibnamefont
  {{Strauss}}}, \bibinfo {author} {\bibfnamefont {M.~R.}\ \bibnamefont
  {{Blanton}}}, \bibinfo {author} {\bibfnamefont {K.}~\bibnamefont
  {{Abazajian}}}, \bibinfo {author} {\bibfnamefont {S.}~\bibnamefont
  {{Dodelson}}}, \bibinfo {author} {\bibfnamefont {H.}~\bibnamefont
  {{Sandvik}}}, \bibinfo {author} {\bibfnamefont {X.}~\bibnamefont {{Wang}}},
  \bibinfo {author} {\bibfnamefont {D.~H.}\ \bibnamefont {{Weinberg}}},
  \bibinfo {author} {\bibfnamefont {I.}~\bibnamefont {{Zehavi}}}, and\ \bibinfo
  {author} {\bibfnamefont {N.~A.}\ \bibnamefont {{Bahcall}}},\ }\bibfield
  {title} {\bibinfo {title} {{Cosmological parameters from SDSS and WMAP}},\
  }\href {https://doi.org/10.1103/PhysRevD.69.103501} {\bibfield  {journal}
  {\bibinfo  {journal} {\prd}\ }\textbf {\bibinfo {volume} {69}},\ \bibinfo
  {eid} {103501} (\bibinfo {year} {2004})},\ \Eprint
  {https://arxiv.org/abs/astro-ph/0310723} {arXiv:astro-ph/0310723 [astro-ph]}
  \BibitemShut {NoStop}%
\bibitem [{\citenamefont {{Rubin}}\ \emph {et~al.}(1980)\citenamefont
  {{Rubin}}, \citenamefont {{Ford}},\ and\ \citenamefont
  {{Thonnard}}}]{Rubin1980}%
  \BibitemOpen
  \bibfield  {author} {\bibinfo {author} {\bibfnamefont {V.~C.}\ \bibnamefont
  {{Rubin}}}, \bibinfo {author} {\bibfnamefont {J.}~\bibnamefont {{Ford}},
  \bibfnamefont {W.~K.}}, and\ \bibinfo {author} {\bibfnamefont
  {N.}~\bibnamefont {{Thonnard}}},\ }\bibfield  {title} {\bibinfo {title}
  {{Rotational properties of 21 SC galaxies with a large range of luminosities
  and radii, from NGC 4605 (R=4kpc) to UGC 2885 (R=122kpc).}},\ }\href
  {https://doi.org/10.1086/158003} {\bibfield  {journal} {\bibinfo  {journal}
  {\apj}\ }\textbf {\bibinfo {volume} {238}},\ \bibinfo {pages} {471} (\bibinfo
  {year} {1980})}\BibitemShut {NoStop}%
\bibitem [{\citenamefont {Akerib}\ \emph {et~al.}(2017)\citenamefont {Akerib}
  \emph {et~al.}}]{Akerib2017}%
  \BibitemOpen
  \bibfield  {author} {\bibinfo {author} {\bibfnamefont {D.~S.}\ \bibnamefont
  {Akerib}} \emph {et~al.} (\bibinfo {collaboration} {LUX Collaboration}),\
  }\bibfield  {title} {\bibinfo {title} {Results from a search for dark matter
  in the complete lux exposure},\ }\href
  {https://doi.org/10.1103/PhysRevLett.118.021303} {\bibfield  {journal}
  {\bibinfo  {journal} {Phys. Rev. Lett.}\ }\textbf {\bibinfo {volume} {118}},\
  \bibinfo {pages} {021303} (\bibinfo {year} {2017})}\BibitemShut {NoStop}%
\bibitem [{\citenamefont {Cui}\ \emph {et~al.}(2017)\citenamefont {Cui} \emph
  {et~al.}}]{Cui2017}%
  \BibitemOpen
  \bibfield  {author} {\bibinfo {author} {\bibfnamefont {X.}~\bibnamefont
  {Cui}} \emph {et~al.} (\bibinfo {collaboration} {PandaX-II Collaboration}),\
  }\bibfield  {title} {\bibinfo {title} {Dark matter results from 54-ton-day
  exposure of pandax-ii experiment},\ }\href
  {https://doi.org/10.1103/PhysRevLett.119.181302} {\bibfield  {journal}
  {\bibinfo  {journal} {Phys. Rev. Lett.}\ }\textbf {\bibinfo {volume} {119}},\
  \bibinfo {pages} {181302} (\bibinfo {year} {2017})}\BibitemShut {NoStop}%
\bibitem [{\citenamefont {{Aprile}}\ \emph {et~al.}(2018)\citenamefont
  {{Aprile}} \emph {et~al.}}]{Aprile2018}%
  \BibitemOpen
  \bibfield  {author} {\bibinfo {author} {\bibfnamefont {E.}~\bibnamefont
  {{Aprile}}} \emph {et~al.} (\bibinfo {collaboration} {Xenon Collaboration}),\
  }\bibfield  {title} {\bibinfo {title} {{Dark Matter Search Results from a One
  Ton-Year Exposure of XENON1T}},\ }\href
  {https://doi.org/10.1103/PhysRevLett.121.111302} {\bibfield  {journal}
  {\bibinfo  {journal} {\prl}\ }\textbf {\bibinfo {volume} {121}},\ \bibinfo
  {eid} {111302} (\bibinfo {year} {2018})},\ \Eprint
  {https://arxiv.org/abs/1805.12562} {arXiv:1805.12562 [astro-ph.CO]}
  \BibitemShut {NoStop}%
\bibitem [{\citenamefont {{Bullock}}\ and\ \citenamefont
  {{Boylan-Kolchin}}(2017)}]{Bullock2017}%
  \BibitemOpen
  \bibfield  {author} {\bibinfo {author} {\bibfnamefont {J.~S.}\ \bibnamefont
  {{Bullock}}}and\ \bibinfo {author} {\bibfnamefont {M.}~\bibnamefont
  {{Boylan-Kolchin}}},\ }\bibfield  {title} {\bibinfo {title} {{Small-Scale
  Challenges to the {\ensuremath{\Lambda}}CDM Paradigm}},\ }\href
  {https://doi.org/10.1146/annurev-astro-091916-055313} {\bibfield  {journal}
  {\bibinfo  {journal} {\araa}\ }\textbf {\bibinfo {volume} {55}},\ \bibinfo
  {pages} {343} (\bibinfo {year} {2017})},\ \Eprint
  {https://arxiv.org/abs/1707.04256} {arXiv:1707.04256 [astro-ph.CO]}
  \BibitemShut {NoStop}%
\bibitem [{\citenamefont {Spergel}\ and\ \citenamefont
  {Steinhardt}(2000)}]{SpergelSteinhardt2000}%
  \BibitemOpen
  \bibfield  {author} {\bibinfo {author} {\bibfnamefont {D.~N.}\ \bibnamefont
  {Spergel}}and\ \bibinfo {author} {\bibfnamefont {P.~J.}\ \bibnamefont
  {Steinhardt}},\ }\bibfield  {title} {\bibinfo {title} {{Observational
  Evidence for Self-Interacting Cold Dark Matter}},\ }\href
  {https://doi.org/10.1103/PhysRevLett.84.3760} {\bibfield  {journal} {\bibinfo
   {journal} {Phys. Rev. Lett.}\ }\textbf {\bibinfo {volume} {84}},\ \bibinfo
  {pages} {3760} (\bibinfo {year} {2000})}\BibitemShut {NoStop}%
\bibitem [{\citenamefont {{Rocha}}\ \emph {et~al.}(2013)\citenamefont
  {{Rocha}}, \citenamefont {{Peter}}, \citenamefont {{Bullock}}, \citenamefont
  {{Kaplinghat}}, \citenamefont {{Garrison-Kimmel}}, \citenamefont
  {{O{\~n}orbe}},\ and\ \citenamefont {{Moustakas}}}]{Rocha2013}%
  \BibitemOpen
  \bibfield  {author} {\bibinfo {author} {\bibfnamefont {M.}~\bibnamefont
  {{Rocha}}}, \bibinfo {author} {\bibfnamefont {A.~H.~G.}\ \bibnamefont
  {{Peter}}}, \bibinfo {author} {\bibfnamefont {J.~S.}\ \bibnamefont
  {{Bullock}}}, \bibinfo {author} {\bibfnamefont {M.}~\bibnamefont
  {{Kaplinghat}}}, \bibinfo {author} {\bibfnamefont {S.}~\bibnamefont
  {{Garrison-Kimmel}}}, \bibinfo {author} {\bibfnamefont {J.}~\bibnamefont
  {{O{\~n}orbe}}}, and\ \bibinfo {author} {\bibfnamefont {L.~A.}\ \bibnamefont
  {{Moustakas}}},\ }\bibfield  {title} {\bibinfo {title} {{Cosmological
  simulations with self-interacting dark matter - I. Constant-density cores and
  substructure}},\ }\href {https://doi.org/10.1093/mnras/sts514} {\bibfield
  {journal} {\bibinfo  {journal} {\mnras}\ }\textbf {\bibinfo {volume} {430}},\
  \bibinfo {pages} {81} (\bibinfo {year} {2013})},\ \Eprint
  {https://arxiv.org/abs/1208.3025} {arXiv:1208.3025 [astro-ph.CO]}
  \BibitemShut {NoStop}%
\bibitem [{\citenamefont {Tulin}\ and\ \citenamefont {Yu}(2018)}]{Tulin2018}%
  \BibitemOpen
  \bibfield  {author} {\bibinfo {author} {\bibfnamefont {S.}~\bibnamefont
  {Tulin}}and\ \bibinfo {author} {\bibfnamefont {H.~B.}\ \bibnamefont {Yu}},\
  }\bibfield  {title} {\bibinfo {title} {{Dark matter self-interactions and
  small scale structure}},\ }\href
  {https://doi.org/10.1016/j.physrep.2017.11.004} {\bibfield  {journal}
  {\bibinfo  {journal} {Phys. Rep.}\ }\textbf {\bibinfo {volume} {730}},\
  \bibinfo {pages} {1} (\bibinfo {year} {2018})}\BibitemShut {NoStop}%
\bibitem [{\citenamefont {Gnedin}\ and\ \citenamefont
  {Ostriker}(2001)}]{Gnedin2001}%
  \BibitemOpen
  \bibfield  {author} {\bibinfo {author} {\bibfnamefont {O.~Y.}\ \bibnamefont
  {Gnedin}}and\ \bibinfo {author} {\bibfnamefont {J.~P.}\ \bibnamefont
  {Ostriker}},\ }\bibfield  {title} {\bibinfo {title} {{Limits on Collisional
  Dark Matter from Elliptical Galaxies in Clusters}},\ }\href
  {https://doi.org/10.1086/323211} {\bibfield  {journal} {\bibinfo  {journal}
  {\apj}\ }\textbf {\bibinfo {volume} {561}},\ \bibinfo {pages} {61} (\bibinfo
  {year} {2001})},\ \Eprint {https://arxiv.org/abs/astro-ph/0010436}
  {arXiv:astro-ph/0010436 [astro-ph]} \BibitemShut {NoStop}%
\bibitem [{\citenamefont {Kahlhoefer}\ \emph {et~al.}(2014)\citenamefont
  {Kahlhoefer}, \citenamefont {Schmidt-Hoberg}, \citenamefont {Frandsen},\ and\
  \citenamefont {Sarkar}}]{Kahlhoefer2014}%
  \BibitemOpen
  \bibfield  {author} {\bibinfo {author} {\bibfnamefont {F.}~\bibnamefont
  {Kahlhoefer}}, \bibinfo {author} {\bibfnamefont {K.}~\bibnamefont
  {Schmidt-Hoberg}}, \bibinfo {author} {\bibfnamefont {M.~T.}\ \bibnamefont
  {Frandsen}}, and\ \bibinfo {author} {\bibfnamefont {S.}~\bibnamefont
  {Sarkar}},\ }\bibfield  {title} {\bibinfo {title} {{Colliding clusters and
  dark matter self-interactions}},\ }\href
  {https://doi.org/10.1093/mnras/stt2097} {\bibfield  {journal} {\bibinfo
  {journal} {Mon. Not. R. Astron. Soc.}\ }\textbf {\bibinfo {volume} {437}},\
  \bibinfo {pages} {2865} (\bibinfo {year} {2014})},\ \Eprint
  {https://arxiv.org/abs/1308.3419} {arXiv:1308.3419} \BibitemShut {NoStop}%
\bibitem [{\citenamefont {{Miralda-Escud{\'e}}}(2002)}]{Miralda-Escude2002}%
  \BibitemOpen
  \bibfield  {author} {\bibinfo {author} {\bibfnamefont {J.}~\bibnamefont
  {{Miralda-Escud{\'e}}}},\ }\bibfield  {title} {\bibinfo {title} {{A Test of
  the Collisional Dark Matter Hypothesis from Cluster Lensing}},\ }\href
  {https://doi.org/10.1086/324138} {\bibfield  {journal} {\bibinfo  {journal}
  {\apj}\ }\textbf {\bibinfo {volume} {564}},\ \bibinfo {pages} {60} (\bibinfo
  {year} {2002})},\ \Eprint {https://arxiv.org/abs/astro-ph/0002050}
  {arXiv:astro-ph/0002050 [astro-ph]} \BibitemShut {NoStop}%
\bibitem [{\citenamefont {{Peter}}\ \emph {et~al.}(2013)\citenamefont
  {{Peter}}, \citenamefont {{Rocha}}, \citenamefont {{Bullock}},\ and\
  \citenamefont {{Kaplinghat}}}]{Peter2013}%
  \BibitemOpen
  \bibfield  {author} {\bibinfo {author} {\bibfnamefont {A.~H.~G.}\
  \bibnamefont {{Peter}}}, \bibinfo {author} {\bibfnamefont {M.}~\bibnamefont
  {{Rocha}}}, \bibinfo {author} {\bibfnamefont {J.~S.}\ \bibnamefont
  {{Bullock}}}, and\ \bibinfo {author} {\bibfnamefont {M.}~\bibnamefont
  {{Kaplinghat}}},\ }\bibfield  {title} {\bibinfo {title} {{Cosmological
  simulations with self-interacting dark matter - II. Halo shapes versus
  observations}},\ }\href {https://doi.org/10.1093/mnras/sts535} {\bibfield
  {journal} {\bibinfo  {journal} {\mnras}\ }\textbf {\bibinfo {volume} {430}},\
  \bibinfo {pages} {105} (\bibinfo {year} {2013})},\ \Eprint
  {https://arxiv.org/abs/1208.3026} {arXiv:1208.3026 [astro-ph.CO]}
  \BibitemShut {NoStop}%
\bibitem [{\citenamefont {Markevitch}\ \emph {et~al.}(2004)\citenamefont
  {Markevitch}, \citenamefont {Gonzalez}, \citenamefont {Clowe}, \citenamefont
  {Vikhlinin}, \citenamefont {Forman}, \citenamefont {Jones}, \citenamefont
  {Murray},\ and\ \citenamefont {Tucker}}]{Markevitch2004}%
  \BibitemOpen
  \bibfield  {author} {\bibinfo {author} {\bibfnamefont {M.}~\bibnamefont
  {Markevitch}}, \bibinfo {author} {\bibfnamefont {A.~H.}\ \bibnamefont
  {Gonzalez}}, \bibinfo {author} {\bibfnamefont {D.}~\bibnamefont {Clowe}},
  \bibinfo {author} {\bibfnamefont {A.}~\bibnamefont {Vikhlinin}}, \bibinfo
  {author} {\bibfnamefont {W.}~\bibnamefont {Forman}}, \bibinfo {author}
  {\bibfnamefont {C.}~\bibnamefont {Jones}}, \bibinfo {author} {\bibfnamefont
  {S.}~\bibnamefont {Murray}}, and\ \bibinfo {author} {\bibfnamefont
  {W.}~\bibnamefont {Tucker}},\ }\bibfield  {title} {\bibinfo {title} {{Direct
  Constraints on the Dark Matter Self--Interaction Cross Section from the
  Merging Galaxy Cluster 1E 0657--56}},\ }\href
  {https://doi.org/10.1086/383178} {\bibfield  {journal} {\bibinfo  {journal}
  {Astrophys. J.}\ }\textbf {\bibinfo {volume} {606}},\ \bibinfo {pages} {819}
  (\bibinfo {year} {2004})},\ \Eprint {https://arxiv.org/abs/0309303}
  {arXiv:0309303 [astro-ph]} \BibitemShut {NoStop}%
\bibitem [{\citenamefont {Randall}\ \emph {et~al.}(2008)\citenamefont
  {Randall}, \citenamefont {Markevitch}, \citenamefont {Clowe}, \citenamefont
  {Gonzalez},\ and\ \citenamefont {Brada{\v{c}}}}]{Randall2008}%
  \BibitemOpen
  \bibfield  {author} {\bibinfo {author} {\bibfnamefont {S.~W.}\ \bibnamefont
  {Randall}}, \bibinfo {author} {\bibfnamefont {M.}~\bibnamefont {Markevitch}},
  \bibinfo {author} {\bibfnamefont {D.}~\bibnamefont {Clowe}}, \bibinfo
  {author} {\bibfnamefont {A.~H.}\ \bibnamefont {Gonzalez}}, and\ \bibinfo
  {author} {\bibfnamefont {M.}~\bibnamefont {Brada{\v{c}}}},\ }\bibfield
  {title} {\bibinfo {title} {{Constraints on the Self‐Interaction Cross
  Section of Dark Matter from Numerical Simulations of the Merging Galaxy
  Cluster 1E 0657--56}},\ }\href {https://doi.org/10.1086/587859} {\bibfield
  {journal} {\bibinfo  {journal} {Astrophys. J.}\ }\textbf {\bibinfo {volume}
  {679}},\ \bibinfo {pages} {1173} (\bibinfo {year} {2008})}\BibitemShut
  {NoStop}%
\bibitem [{\citenamefont {{Robertson}}\ \emph {et~al.}(2017)\citenamefont
  {{Robertson}}, \citenamefont {{Massey}},\ and\ \citenamefont
  {{Eke}}}]{Robertson2017}%
  \BibitemOpen
  \bibfield  {author} {\bibinfo {author} {\bibfnamefont {A.}~\bibnamefont
  {{Robertson}}}, \bibinfo {author} {\bibfnamefont {R.}~\bibnamefont
  {{Massey}}}, and\ \bibinfo {author} {\bibfnamefont {V.}~\bibnamefont
  {{Eke}}},\ }\bibfield  {title} {\bibinfo {title} {{What does the Bullet
  Cluster tell us about self-interacting dark matter?}},\ }\href
  {https://doi.org/10.1093/mnras/stw2670} {\bibfield  {journal} {\bibinfo
  {journal} {\mnras}\ }\textbf {\bibinfo {volume} {465}},\ \bibinfo {pages}
  {569} (\bibinfo {year} {2017})},\ \Eprint {https://arxiv.org/abs/1605.04307}
  {arXiv:1605.04307 [astro-ph.CO]} \BibitemShut {NoStop}%
\bibitem [{\citenamefont {Massey}\ \emph {et~al.}(2011)\citenamefont {Massey},
  \citenamefont {Kitching},\ and\ \citenamefont {Nagai}}]{Massey2011}%
  \BibitemOpen
  \bibfield  {author} {\bibinfo {author} {\bibfnamefont {R.}~\bibnamefont
  {Massey}}, \bibinfo {author} {\bibfnamefont {T.}~\bibnamefont {Kitching}},
  and\ \bibinfo {author} {\bibfnamefont {D.}~\bibnamefont {Nagai}},\ }\bibfield
   {title} {\bibinfo {title} {{Cluster bulleticity}},\ }\href
  {https://doi.org/10.1111/j.1365-2966.2011.18246.x} {\bibfield  {journal}
  {\bibinfo  {journal} {Mon. Not. R. Astron. Soc.}\ }\textbf {\bibinfo {volume}
  {413}},\ \bibinfo {pages} {1709} (\bibinfo {year} {2011})}\BibitemShut
  {NoStop}%
\bibitem [{\citenamefont {Harvey}\ \emph {et~al.}(2013)\citenamefont {Harvey},
  \citenamefont {Massey}, \citenamefont {Kitching}, \citenamefont {Taylor},
  \citenamefont {Jullo}, \citenamefont {Kneib}, \citenamefont {Tittley},\ and\
  \citenamefont {Marshall}}]{Harvey2013}%
  \BibitemOpen
  \bibfield  {author} {\bibinfo {author} {\bibfnamefont {D.}~\bibnamefont
  {Harvey}}, \bibinfo {author} {\bibfnamefont {R.}~\bibnamefont {Massey}},
  \bibinfo {author} {\bibfnamefont {T.}~\bibnamefont {Kitching}}, \bibinfo
  {author} {\bibfnamefont {A.}~\bibnamefont {Taylor}}, \bibinfo {author}
  {\bibfnamefont {E.}~\bibnamefont {Jullo}}, \bibinfo {author} {\bibfnamefont
  {J.~P.}\ \bibnamefont {Kneib}}, \bibinfo {author} {\bibfnamefont
  {E.}~\bibnamefont {Tittley}}, and\ \bibinfo {author} {\bibfnamefont {P.~J.}\
  \bibnamefont {Marshall}},\ }\bibfield  {title} {\bibinfo {title} {{Dark
  matter astrometry: Accuracy of subhalo positions for the measurement of
  self-interaction cross-sections}},\ }\href
  {https://doi.org/10.1093/mnras/stt819} {\bibfield  {journal} {\bibinfo
  {journal} {Mon. Not. R. Astron. Soc.}\ }\textbf {\bibinfo {volume} {433}},\
  \bibinfo {pages} {1517} (\bibinfo {year} {2013})},\ \Eprint
  {https://arxiv.org/abs/1305.2117} {arXiv:1305.2117} \BibitemShut {NoStop}%
\bibitem [{\citenamefont {{Banerjee}}\ \emph {et~al.}(2019)\citenamefont
  {{Banerjee}}, \citenamefont {{Adhikari}}, \citenamefont {{Dalal}},
  \citenamefont {{More}},\ and\ \citenamefont {{Kravtsov}}}]{Banerjee2019}%
  \BibitemOpen
  \bibfield  {author} {\bibinfo {author} {\bibfnamefont {A.}~\bibnamefont
  {{Banerjee}}}, \bibinfo {author} {\bibfnamefont {S.}~\bibnamefont
  {{Adhikari}}}, \bibinfo {author} {\bibfnamefont {N.}~\bibnamefont {{Dalal}}},
  \bibinfo {author} {\bibfnamefont {S.}~\bibnamefont {{More}}}, and\ \bibinfo
  {author} {\bibfnamefont {A.}~\bibnamefont {{Kravtsov}}},\ }\bibfield  {title}
  {\bibinfo {title} {{Signatures of Self-Interacting dark matter on cluster
  density profile and subhalo distributions}},\ }\href@noop {} {\bibfield
  {journal} {\bibinfo  {journal} {arXiv e-prints}\ ,\ \bibinfo {eid}
  {arXiv:1906.12026}} (\bibinfo {year} {2019})},\ \Eprint
  {https://arxiv.org/abs/1906.12026} {arXiv:1906.12026 [astro-ph.CO]}
  \BibitemShut {NoStop}%
\bibitem [{\citenamefont {Secco}\ \emph {et~al.}(2018)\citenamefont {Secco},
  \citenamefont {Farah}, \citenamefont {Jain}, \citenamefont {Adhikari},
  \citenamefont {Banerjee},\ and\ \citenamefont {Dalal}}]{Secco2017}%
  \BibitemOpen
  \bibfield  {author} {\bibinfo {author} {\bibfnamefont {L.~F.}\ \bibnamefont
  {Secco}}, \bibinfo {author} {\bibfnamefont {A.}~\bibnamefont {Farah}},
  \bibinfo {author} {\bibfnamefont {B.}~\bibnamefont {Jain}}, \bibinfo {author}
  {\bibfnamefont {S.}~\bibnamefont {Adhikari}}, \bibinfo {author}
  {\bibfnamefont {A.}~\bibnamefont {Banerjee}}, and\ \bibinfo {author}
  {\bibfnamefont {N.}~\bibnamefont {Dalal}},\ }\bibfield  {title} {\bibinfo
  {title} {{Probing Self-interacting Dark Matter with Disk Galaxies in Cluster
  Environments}},\ }\href {https://doi.org/10.3847/1538-4357/aac271} {\bibfield
   {journal} {\bibinfo  {journal} {Astrophys. J.}\ }\textbf {\bibinfo {volume}
  {860}},\ \bibinfo {pages} {32} (\bibinfo {year} {2018})},\ \Eprint
  {https://arxiv.org/abs/1712.04841} {arXiv:1712.04841} \BibitemShut {NoStop}%
\bibitem [{\citenamefont {Kummer}\ \emph {et~al.}(2018)\citenamefont {Kummer},
  \citenamefont {Kahlhoefer},\ and\ \citenamefont
  {Schmidt-Hoberg}}]{Kummer2017}%
  \BibitemOpen
  \bibfield  {author} {\bibinfo {author} {\bibfnamefont {J.}~\bibnamefont
  {Kummer}}, \bibinfo {author} {\bibfnamefont {F.}~\bibnamefont {Kahlhoefer}},
  and\ \bibinfo {author} {\bibfnamefont {K.}~\bibnamefont {Schmidt-Hoberg}},\
  }\bibfield  {title} {\bibinfo {title} {{Effective description of dark matter
  self-interactions in small dark matter haloes}},\ }\href
  {https://doi.org/10.1093/mnras/stx2715} {\bibfield  {journal} {\bibinfo
  {journal} {Mon. Not. R. Astron. Soc.}\ }\textbf {\bibinfo {volume} {474}},\
  \bibinfo {pages} {388} (\bibinfo {year} {2018})},\ \Eprint
  {https://arxiv.org/abs/1706.04794} {arXiv:1706.04794} \BibitemShut {NoStop}%
\bibitem [{\citenamefont {{Desmond}}\ \emph
  {et~al.}(2018{\natexlab{a}})\citenamefont {{Desmond}}, \citenamefont
  {{Ferreira}}, \citenamefont {{Lavaux}},\ and\ \citenamefont
  {{Jasche}}}]{Desmond2018b}%
  \BibitemOpen
  \bibfield  {author} {\bibinfo {author} {\bibfnamefont {H.}~\bibnamefont
  {{Desmond}}}, \bibinfo {author} {\bibfnamefont {P.~G.}\ \bibnamefont
  {{Ferreira}}}, \bibinfo {author} {\bibfnamefont {G.}~\bibnamefont
  {{Lavaux}}}, and\ \bibinfo {author} {\bibfnamefont {J.}~\bibnamefont
  {{Jasche}}},\ }\bibfield  {title} {\bibinfo {title} {{Fifth force constraints
  from galaxy warps}},\ }\href {https://doi.org/10.1103/PhysRevD.98.083010}
  {\bibfield  {journal} {\bibinfo  {journal} {\prd}\ }\textbf {\bibinfo
  {volume} {98}},\ \bibinfo {eid} {083010} (\bibinfo {year}
  {2018}{\natexlab{a}})},\ \Eprint {https://arxiv.org/abs/1807.11742}
  {arXiv:1807.11742 [astro-ph.CO]} \BibitemShut {NoStop}%
\bibitem [{\citenamefont {{Vikram}}\ \emph {et~al.}(2013)\citenamefont
  {{Vikram}}, \citenamefont {{Cabr{\'e}}}, \citenamefont {{Jain}},\ and\
  \citenamefont {{Vand erPlas}}}]{Vikram2013}%
  \BibitemOpen
  \bibfield  {author} {\bibinfo {author} {\bibfnamefont {V.}~\bibnamefont
  {{Vikram}}}, \bibinfo {author} {\bibfnamefont {A.}~\bibnamefont
  {{Cabr{\'e}}}}, \bibinfo {author} {\bibfnamefont {B.}~\bibnamefont {{Jain}}},
  and\ \bibinfo {author} {\bibfnamefont {J.~T.}\ \bibnamefont {{Vand
  erPlas}}},\ }\bibfield  {title} {\bibinfo {title} {{Astrophysical tests of
  modified gravity: the morphology and kinematics of dwarf galaxies}},\ }\href
  {https://doi.org/10.1088/1475-7516/2013/08/020} {\bibfield  {journal}
  {\bibinfo  {journal} {\jcap}\ }\textbf {\bibinfo {volume} {2013}},\ \bibinfo
  {eid} {020} (\bibinfo {year} {2013})},\ \Eprint
  {https://arxiv.org/abs/1303.0295} {arXiv:1303.0295 [astro-ph.CO]}
  \BibitemShut {NoStop}%
\bibitem [{\citenamefont {Blanton}\ \emph {et~al.}(2011)\citenamefont
  {Blanton}, \citenamefont {Kazin}, \citenamefont {Muna}, \citenamefont
  {Weaver},\ and\ \citenamefont {Price-Whelan}}]{Blanton2011}%
  \BibitemOpen
  \bibfield  {author} {\bibinfo {author} {\bibfnamefont {M.~R.}\ \bibnamefont
  {Blanton}}, \bibinfo {author} {\bibfnamefont {E.}~\bibnamefont {Kazin}},
  \bibinfo {author} {\bibfnamefont {D.}~\bibnamefont {Muna}}, \bibinfo {author}
  {\bibfnamefont {B.~A.}\ \bibnamefont {Weaver}}, and\ \bibinfo {author}
  {\bibfnamefont {A.}~\bibnamefont {Price-Whelan}},\ }\bibfield  {title}
  {\bibinfo {title} {{Improved Background Subtraction for the Sloan Digital Sky
  Survey Images}},\ }\href {https://doi.org/10.1088/0004-6256/142/1/31}
  {\bibfield  {journal} {\bibinfo  {journal} {\aj}\ }\textbf {\bibinfo {volume}
  {142}},\ \bibinfo {pages} {31} (\bibinfo {year} {2011})},\ \Eprint
  {https://arxiv.org/abs/1105.1960} {arXiv:1105.1960 [astro-ph.IM]}
  \BibitemShut {NoStop}%
\bibitem [{\citenamefont {{Kravtsov}}\ \emph {et~al.}(2004)\citenamefont
  {{Kravtsov}}, \citenamefont {{Berlind}}, \citenamefont {{Wechsler}},
  \citenamefont {{Klypin}}, \citenamefont {{Gottl{\"o}ber}}, \citenamefont
  {{Allgood}},\ and\ \citenamefont {{Primack}}}]{Kravtsov2004}%
  \BibitemOpen
  \bibfield  {author} {\bibinfo {author} {\bibfnamefont {A.~V.}\ \bibnamefont
  {{Kravtsov}}}, \bibinfo {author} {\bibfnamefont {A.~A.}\ \bibnamefont
  {{Berlind}}}, \bibinfo {author} {\bibfnamefont {R.~H.}\ \bibnamefont
  {{Wechsler}}}, \bibinfo {author} {\bibfnamefont {A.~A.}\ \bibnamefont
  {{Klypin}}}, \bibinfo {author} {\bibfnamefont {S.}~\bibnamefont
  {{Gottl{\"o}ber}}}, \bibinfo {author} {\bibfnamefont {B.~o.}\ \bibnamefont
  {{Allgood}}}, and\ \bibinfo {author} {\bibfnamefont {J.~R.}\ \bibnamefont
  {{Primack}}},\ }\bibfield  {title} {\bibinfo {title} {{The Dark Side of the
  Halo Occupation Distribution}},\ }\href {https://doi.org/10.1086/420959}
  {\bibfield  {journal} {\bibinfo  {journal} {\apj}\ }\textbf {\bibinfo
  {volume} {609}},\ \bibinfo {pages} {35} (\bibinfo {year} {2004})},\ \Eprint
  {https://arxiv.org/abs/astro-ph/0308519} {arXiv:astro-ph/0308519 [astro-ph]}
  \BibitemShut {NoStop}%
\bibitem [{\citenamefont {{Lehmann}}\ \emph {et~al.}(2017)\citenamefont
  {{Lehmann}}, \citenamefont {{Mao}}, \citenamefont {{Becker}}, \citenamefont
  {{Skillman}},\ and\ \citenamefont {{Wechsler}}}]{Lehmann2017}%
  \BibitemOpen
  \bibfield  {author} {\bibinfo {author} {\bibfnamefont {B.~V.}\ \bibnamefont
  {{Lehmann}}}, \bibinfo {author} {\bibfnamefont {Y.-Y.}\ \bibnamefont
  {{Mao}}}, \bibinfo {author} {\bibfnamefont {M.~R.}\ \bibnamefont {{Becker}}},
  \bibinfo {author} {\bibfnamefont {S.~W.}\ \bibnamefont {{Skillman}}}, and\
  \bibinfo {author} {\bibfnamefont {R.~H.}\ \bibnamefont {{Wechsler}}},\
  }\bibfield  {title} {\bibinfo {title} {{The Concentration Dependence of the
  Galaxy-Halo Connection: Modeling Assembly Bias with Abundance Matching}},\
  }\href {https://doi.org/10.3847/1538-4357/834/1/37} {\bibfield  {journal}
  {\bibinfo  {journal} {\apj}\ }\textbf {\bibinfo {volume} {834}},\ \bibinfo
  {eid} {37} (\bibinfo {year} {2017})},\ \Eprint
  {https://arxiv.org/abs/1510.05651} {arXiv:1510.05651 [astro-ph.CO]}
  \BibitemShut {NoStop}%
\bibitem [{\citenamefont {{Skillman}}\ \emph {et~al.}(2014)\citenamefont
  {{Skillman}}, \citenamefont {{Warren}}, \citenamefont {{Turk}}, \citenamefont
  {{Wechsler}}, \citenamefont {{Holz}},\ and\ \citenamefont
  {{Sutter}}}]{Skillman2014}%
  \BibitemOpen
  \bibfield  {author} {\bibinfo {author} {\bibfnamefont {S.~W.}\ \bibnamefont
  {{Skillman}}}, \bibinfo {author} {\bibfnamefont {M.~S.}\ \bibnamefont
  {{Warren}}}, \bibinfo {author} {\bibfnamefont {M.~J.}\ \bibnamefont
  {{Turk}}}, \bibinfo {author} {\bibfnamefont {R.~H.}\ \bibnamefont
  {{Wechsler}}}, \bibinfo {author} {\bibfnamefont {D.~E.}\ \bibnamefont
  {{Holz}}}, and\ \bibinfo {author} {\bibfnamefont {P.~M.}\ \bibnamefont
  {{Sutter}}},\ }\bibfield  {title} {\bibinfo {title} {{Dark Sky Simulations:
  Early Data Release}},\ }\href@noop {} {\bibfield  {journal} {\bibinfo
  {journal} {arXiv e-prints}\ ,\ \bibinfo {eid} {arXiv:1407.2600}} (\bibinfo
  {year} {2014})},\ \Eprint {https://arxiv.org/abs/1407.2600} {arXiv:1407.2600
  [astro-ph.CO]} \BibitemShut {NoStop}%
\bibitem [{\citenamefont {{Behroozi}}\ \emph {et~al.}(2013)\citenamefont
  {{Behroozi}}, \citenamefont {{Wechsler}},\ and\ \citenamefont
  {{Wu}}}]{Behroozi2013}%
  \BibitemOpen
  \bibfield  {author} {\bibinfo {author} {\bibfnamefont {P.~S.}\ \bibnamefont
  {{Behroozi}}}, \bibinfo {author} {\bibfnamefont {R.~H.}\ \bibnamefont
  {{Wechsler}}}, and\ \bibinfo {author} {\bibfnamefont {H.-Y.}\ \bibnamefont
  {{Wu}}},\ }\bibfield  {title} {\bibinfo {title} {{The ROCKSTAR Phase-space
  Temporal Halo Finder and the Velocity Offsets of Cluster Cores}},\ }\href
  {https://doi.org/10.1088/0004-637X/762/2/109} {\bibfield  {journal} {\bibinfo
   {journal} {\apj}\ }\textbf {\bibinfo {volume} {762}},\ \bibinfo {eid} {109}
  (\bibinfo {year} {2013})},\ \Eprint {https://arxiv.org/abs/1110.4372}
  {arXiv:1110.4372 [astro-ph.CO]} \BibitemShut {NoStop}%
\bibitem [{\citenamefont {{Navarro}}\ \emph {et~al.}(1997)\citenamefont
  {{Navarro}}, \citenamefont {{Frenk}},\ and\ \citenamefont
  {{White}}}]{Navarro1997}%
  \BibitemOpen
  \bibfield  {author} {\bibinfo {author} {\bibfnamefont {J.~F.}\ \bibnamefont
  {{Navarro}}}, \bibinfo {author} {\bibfnamefont {C.~S.}\ \bibnamefont
  {{Frenk}}}, and\ \bibinfo {author} {\bibfnamefont {S.~D.~M.}\ \bibnamefont
  {{White}}},\ }\bibfield  {title} {\bibinfo {title} {{A Universal Density
  Profile from Hierarchical Clustering}},\ }\href
  {https://doi.org/10.1086/304888} {\bibfield  {journal} {\bibinfo  {journal}
  {\apj}\ }\textbf {\bibinfo {volume} {490}},\ \bibinfo {pages} {493} (\bibinfo
  {year} {1997})},\ \Eprint {https://arxiv.org/abs/astro-ph/9611107}
  {arXiv:astro-ph/9611107 [astro-ph]} \BibitemShut {NoStop}%
\bibitem [{\citenamefont {{Jasche}}\ \emph {et~al.}(2010)\citenamefont
  {{Jasche}}, \citenamefont {{Kitaura}}, \citenamefont {{Wandelt}},\ and\
  \citenamefont {{En{\ss}lin}}}]{Jasche2010}%
  \BibitemOpen
  \bibfield  {author} {\bibinfo {author} {\bibfnamefont {J.}~\bibnamefont
  {{Jasche}}}, \bibinfo {author} {\bibfnamefont {F.~S.}\ \bibnamefont
  {{Kitaura}}}, \bibinfo {author} {\bibfnamefont {B.~D.}\ \bibnamefont
  {{Wandelt}}}, and\ \bibinfo {author} {\bibfnamefont {T.~A.}\ \bibnamefont
  {{En{\ss}lin}}},\ }\bibfield  {title} {\bibinfo {title} {{Bayesian
  power-spectrum inference for large-scale structure data}},\ }\href
  {https://doi.org/10.1111/j.1365-2966.2010.16610.x} {\bibfield  {journal}
  {\bibinfo  {journal} {\mnras}\ }\textbf {\bibinfo {volume} {406}},\ \bibinfo
  {pages} {60} (\bibinfo {year} {2010})},\ \Eprint
  {https://arxiv.org/abs/0911.2493} {arXiv:0911.2493 [astro-ph.CO]}
  \BibitemShut {NoStop}%
\bibitem [{\citenamefont {{Jasche}}\ and\ \citenamefont
  {{Wandelt}}(2012)}]{Jasche2012}%
  \BibitemOpen
  \bibfield  {author} {\bibinfo {author} {\bibfnamefont {J.}~\bibnamefont
  {{Jasche}}}and\ \bibinfo {author} {\bibfnamefont {B.~D.}\ \bibnamefont
  {{Wandelt}}},\ }\bibfield  {title} {\bibinfo {title} {{Bayesian inference
  from photometric redshift surveys}},\ }\href
  {https://doi.org/10.1111/j.1365-2966.2012.21423.x} {\bibfield  {journal}
  {\bibinfo  {journal} {\mnras}\ }\textbf {\bibinfo {volume} {425}},\ \bibinfo
  {pages} {1042} (\bibinfo {year} {2012})},\ \Eprint
  {https://arxiv.org/abs/1106.2757} {arXiv:1106.2757 [astro-ph.CO]}
  \BibitemShut {NoStop}%
\bibitem [{\citenamefont {{Jasche}}\ and\ \citenamefont
  {{Wandelt}}(2013)}]{Jasche2013}%
  \BibitemOpen
  \bibfield  {author} {\bibinfo {author} {\bibfnamefont {J.}~\bibnamefont
  {{Jasche}}}and\ \bibinfo {author} {\bibfnamefont {B.~D.}\ \bibnamefont
  {{Wandelt}}},\ }\bibfield  {title} {\bibinfo {title} {{Bayesian physical
  reconstruction of initial conditions from large-scale structure surveys}},\
  }\href {https://doi.org/10.1093/mnras/stt449} {\bibfield  {journal} {\bibinfo
   {journal} {\mnras}\ }\textbf {\bibinfo {volume} {432}},\ \bibinfo {pages}
  {894} (\bibinfo {year} {2013})},\ \Eprint {https://arxiv.org/abs/1203.3639}
  {arXiv:1203.3639 [astro-ph.CO]} \BibitemShut {NoStop}%
\bibitem [{\citenamefont {{Jasche}}\ \emph {et~al.}(2015)\citenamefont
  {{Jasche}}, \citenamefont {{Leclercq}},\ and\ \citenamefont
  {{Wandelt}}}]{Jasche2015a}%
  \BibitemOpen
  \bibfield  {author} {\bibinfo {author} {\bibfnamefont {J.}~\bibnamefont
  {{Jasche}}}, \bibinfo {author} {\bibfnamefont {F.}~\bibnamefont
  {{Leclercq}}}, and\ \bibinfo {author} {\bibfnamefont {B.~D.}\ \bibnamefont
  {{Wandelt}}},\ }\bibfield  {title} {\bibinfo {title} {{Past and present
  cosmic structure in the SDSS DR7 main sample}},\ }\href
  {https://doi.org/10.1088/1475-7516/2015/01/036} {\bibfield  {journal}
  {\bibinfo  {journal} {\jcap}\ }\textbf {\bibinfo {volume} {2015}},\ \bibinfo
  {eid} {036} (\bibinfo {year} {2015})},\ \Eprint
  {https://arxiv.org/abs/1409.6308} {arXiv:1409.6308 [astro-ph.CO]}
  \BibitemShut {NoStop}%
\bibitem [{\citenamefont {{Jasche}}\ and\ \citenamefont
  {{Lavaux}}(2015)}]{Jasche2015b}%
  \BibitemOpen
  \bibfield  {author} {\bibinfo {author} {\bibfnamefont {J.}~\bibnamefont
  {{Jasche}}}and\ \bibinfo {author} {\bibfnamefont {G.}~\bibnamefont
  {{Lavaux}}},\ }\bibfield  {title} {\bibinfo {title} {{Matrix-free large-scale
  Bayesian inference in cosmology}},\ }\href
  {https://doi.org/10.1093/mnras/stu2479} {\bibfield  {journal} {\bibinfo
  {journal} {\mnras}\ }\textbf {\bibinfo {volume} {447}},\ \bibinfo {pages}
  {1204} (\bibinfo {year} {2015})},\ \Eprint {https://arxiv.org/abs/1402.1763}
  {arXiv:1402.1763 [astro-ph.CO]} \BibitemShut {NoStop}%
\bibitem [{\citenamefont {{Lavaux}}\ and\ \citenamefont
  {{Jasche}}(2016)}]{Lavaux2016}%
  \BibitemOpen
  \bibfield  {author} {\bibinfo {author} {\bibfnamefont {G.}~\bibnamefont
  {{Lavaux}}}and\ \bibinfo {author} {\bibfnamefont {J.}~\bibnamefont
  {{Jasche}}},\ }\bibfield  {title} {\bibinfo {title} {{Unmasking the masked
  Universe: the 2M++ catalogue through Bayesian eyes}},\ }\href
  {https://doi.org/10.1093/mnras/stv2499} {\bibfield  {journal} {\bibinfo
  {journal} {\mnras}\ }\textbf {\bibinfo {volume} {455}},\ \bibinfo {pages}
  {3169} (\bibinfo {year} {2016})},\ \Eprint {https://arxiv.org/abs/1509.05040}
  {arXiv:1509.05040 [astro-ph.CO]} \BibitemShut {NoStop}%
\bibitem [{\citenamefont {Jasche}\ and\ \citenamefont {Lavaux}(2019)}]{BorgPM}%
  \BibitemOpen
  \bibfield  {author} {\bibinfo {author} {\bibfnamefont {J.}~\bibnamefont
  {Jasche}}and\ \bibinfo {author} {\bibfnamefont {G.}~\bibnamefont {Lavaux}},\
  }\bibfield  {title} {\bibinfo {title} {{Physical Bayesian modelling of the
  non-linear matter distribution: new insights into the Nearby Universe}},\
  }\href {https://doi.org/10.1051/0004-6361/201833710} {\bibfield  {journal}
  {\bibinfo  {journal} {Astron. Astrophys.}\ }\textbf {\bibinfo {volume}
  {625}},\ \bibinfo {pages} {A64} (\bibinfo {year} {2019})},\ \Eprint
  {https://arxiv.org/abs/1806.11117} {arXiv:1806.11117 [astro-ph.CO]}
  \BibitemShut {NoStop}%
\bibitem [{\citenamefont {{Lavaux}}\ and\ \citenamefont
  {{Hudson}}(2011)}]{Lavaux2011}%
  \BibitemOpen
  \bibfield  {author} {\bibinfo {author} {\bibfnamefont {G.}~\bibnamefont
  {{Lavaux}}}and\ \bibinfo {author} {\bibfnamefont {M.~J.}\ \bibnamefont
  {{Hudson}}},\ }\bibfield  {title} {\bibinfo {title} {{The 2M++ galaxy
  redshift catalogue}},\ }\href
  {https://doi.org/10.1111/j.1365-2966.2011.19233.x} {\bibfield  {journal}
  {\bibinfo  {journal} {\mnras}\ }\textbf {\bibinfo {volume} {416}},\ \bibinfo
  {pages} {2840} (\bibinfo {year} {2011})},\ \Eprint
  {https://arxiv.org/abs/1105.6107} {arXiv:1105.6107 [astro-ph.CO]}
  \BibitemShut {NoStop}%
\bibitem [{\citenamefont {{Desmond}}\ \emph
  {et~al.}(2018{\natexlab{b}})\citenamefont {{Desmond}}, \citenamefont
  {{Ferreira}}, \citenamefont {{Lavaux}},\ and\ \citenamefont
  {{Jasche}}}]{Desmond2018maps}%
  \BibitemOpen
  \bibfield  {author} {\bibinfo {author} {\bibfnamefont {H.}~\bibnamefont
  {{Desmond}}}, \bibinfo {author} {\bibfnamefont {P.~G.}\ \bibnamefont
  {{Ferreira}}}, \bibinfo {author} {\bibfnamefont {G.}~\bibnamefont
  {{Lavaux}}}, and\ \bibinfo {author} {\bibfnamefont {J.}~\bibnamefont
  {{Jasche}}},\ }\bibfield  {title} {\bibinfo {title} {{Reconstructing the
  gravitational field of the local Universe}},\ }\href
  {https://doi.org/10.1093/mnras/stx3062} {\bibfield  {journal} {\bibinfo
  {journal} {\mnras}\ }\textbf {\bibinfo {volume} {474}},\ \bibinfo {pages}
  {3152} (\bibinfo {year} {2018}{\natexlab{b}})},\ \Eprint
  {https://arxiv.org/abs/1705.02420} {arXiv:1705.02420 [astro-ph.GA]}
  \BibitemShut {NoStop}%
\bibitem [{\citenamefont {{Tully}}\ \emph {et~al.}(2016)\citenamefont
  {{Tully}}, \citenamefont {{Courtois}},\ and\ \citenamefont
  {{Sorce}}}]{Tully2016}%
  \BibitemOpen
  \bibfield  {author} {\bibinfo {author} {\bibfnamefont {R.~B.}\ \bibnamefont
  {{Tully}}}, \bibinfo {author} {\bibfnamefont {H.~M.}\ \bibnamefont
  {{Courtois}}}, and\ \bibinfo {author} {\bibfnamefont {J.~G.}\ \bibnamefont
  {{Sorce}}},\ }\bibfield  {title} {\bibinfo {title} {{Cosmicflows-3}},\ }\href
  {https://doi.org/10.3847/0004-6256/152/2/50} {\bibfield  {journal} {\bibinfo
  {journal} {\aj}\ }\textbf {\bibinfo {volume} {152}},\ \bibinfo {eid} {50}
  (\bibinfo {year} {2016})},\ \Eprint {https://arxiv.org/abs/1605.01765}
  {arXiv:1605.01765 [astro-ph.CO]} \BibitemShut {NoStop}%
\bibitem [{\citenamefont {{Carrick}}\ \emph {et~al.}(2015)\citenamefont
  {{Carrick}}, \citenamefont {{Turnbull}}, \citenamefont {{Lavaux}},\ and\
  \citenamefont {{Hudson}}}]{Carrick2015}%
  \BibitemOpen
  \bibfield  {author} {\bibinfo {author} {\bibfnamefont {J.}~\bibnamefont
  {{Carrick}}}, \bibinfo {author} {\bibfnamefont {S.~J.}\ \bibnamefont
  {{Turnbull}}}, \bibinfo {author} {\bibfnamefont {G.}~\bibnamefont
  {{Lavaux}}}, and\ \bibinfo {author} {\bibfnamefont {M.~J.}\ \bibnamefont
  {{Hudson}}},\ }\bibfield  {title} {\bibinfo {title} {{Cosmological parameters
  from the comparison of peculiar velocities with predictions from the 2M++
  density field}},\ }\href {https://doi.org/10.1093/mnras/stv547} {\bibfield
  {journal} {\bibinfo  {journal} {\mnras}\ }\textbf {\bibinfo {volume} {450}},\
  \bibinfo {pages} {317} (\bibinfo {year} {2015})},\ \Eprint
  {https://arxiv.org/abs/1504.04627} {arXiv:1504.04627 [astro-ph.CO]}
  \BibitemShut {NoStop}%
\bibitem [{\citenamefont {{Dubois}}\ \emph {et~al.}(2014)\citenamefont
  {{Dubois}}, \citenamefont {{Pichon}}, \citenamefont {{Welker}}, \citenamefont
  {{Le Borgne}}, \citenamefont {{Devriendt}}, \citenamefont {{Laigle}},
  \citenamefont {{Codis}}, \citenamefont {{Pogosyan}}, \citenamefont
  {{Arnouts}}, \citenamefont {{Benabed}}, \citenamefont {{Bertin}},
  \citenamefont {{Blaizot}}, \citenamefont {{Bouchet}}, \citenamefont
  {{Cardoso}}, \citenamefont {{Colombi}}, \citenamefont {{de Lapparent}},
  \citenamefont {{Desjacques}}, \citenamefont {{Gavazzi}}, \citenamefont
  {{Kassin}}, \citenamefont {{Kimm}}, \citenamefont {{McCracken}},
  \citenamefont {{Milliard}}, \citenamefont {{Peirani}}, \citenamefont
  {{Prunet}}, \citenamefont {{Rouberol}}, \citenamefont {{Silk}}, \citenamefont
  {{Slyz}}, \citenamefont {{Sousbie}}, \citenamefont {{Teyssier}},
  \citenamefont {{Tresse}}, \citenamefont {{Treyer}}, \citenamefont
  {{Vibert}},\ and\ \citenamefont {{Volonteri}}}]{Dubois2014}%
  \BibitemOpen
  \bibfield  {author} {\bibinfo {author} {\bibfnamefont {Y.}~\bibnamefont
  {{Dubois}}}, \bibinfo {author} {\bibfnamefont {C.}~\bibnamefont {{Pichon}}},
  \bibinfo {author} {\bibfnamefont {C.}~\bibnamefont {{Welker}}}, \bibinfo
  {author} {\bibfnamefont {D.}~\bibnamefont {{Le Borgne}}}, \bibinfo {author}
  {\bibfnamefont {J.}~\bibnamefont {{Devriendt}}}, \bibinfo {author}
  {\bibfnamefont {C.}~\bibnamefont {{Laigle}}}, \bibinfo {author}
  {\bibfnamefont {S.}~\bibnamefont {{Codis}}}, \bibinfo {author} {\bibfnamefont
  {D.}~\bibnamefont {{Pogosyan}}}, \bibinfo {author} {\bibfnamefont
  {S.}~\bibnamefont {{Arnouts}}}, \bibinfo {author} {\bibfnamefont
  {K.}~\bibnamefont {{Benabed}}}, \bibinfo {author} {\bibfnamefont
  {E.}~\bibnamefont {{Bertin}}}, \bibinfo {author} {\bibfnamefont
  {J.}~\bibnamefont {{Blaizot}}}, \bibinfo {author} {\bibfnamefont
  {F.}~\bibnamefont {{Bouchet}}}, \bibinfo {author} {\bibfnamefont {J.~F.}\
  \bibnamefont {{Cardoso}}}, \bibinfo {author} {\bibfnamefont {S.}~\bibnamefont
  {{Colombi}}}, \bibinfo {author} {\bibfnamefont {V.}~\bibnamefont {{de
  Lapparent}}}, \bibinfo {author} {\bibfnamefont {V.}~\bibnamefont
  {{Desjacques}}}, \bibinfo {author} {\bibfnamefont {R.}~\bibnamefont
  {{Gavazzi}}}, \bibinfo {author} {\bibfnamefont {S.}~\bibnamefont {{Kassin}}},
  \bibinfo {author} {\bibfnamefont {T.}~\bibnamefont {{Kimm}}}, \bibinfo
  {author} {\bibfnamefont {H.}~\bibnamefont {{McCracken}}}, \bibinfo {author}
  {\bibfnamefont {B.}~\bibnamefont {{Milliard}}}, \bibinfo {author}
  {\bibfnamefont {S.}~\bibnamefont {{Peirani}}}, \bibinfo {author}
  {\bibfnamefont {S.}~\bibnamefont {{Prunet}}}, \bibinfo {author}
  {\bibfnamefont {S.}~\bibnamefont {{Rouberol}}}, \bibinfo {author}
  {\bibfnamefont {J.}~\bibnamefont {{Silk}}}, \bibinfo {author} {\bibfnamefont
  {A.}~\bibnamefont {{Slyz}}}, \bibinfo {author} {\bibfnamefont
  {T.}~\bibnamefont {{Sousbie}}}, \bibinfo {author} {\bibfnamefont
  {R.}~\bibnamefont {{Teyssier}}}, \bibinfo {author} {\bibfnamefont
  {L.}~\bibnamefont {{Tresse}}}, \bibinfo {author} {\bibfnamefont
  {M.}~\bibnamefont {{Treyer}}}, \bibinfo {author} {\bibfnamefont
  {D.}~\bibnamefont {{Vibert}}}, and\ \bibinfo {author} {\bibfnamefont
  {M.}~\bibnamefont {{Volonteri}}},\ }\bibfield  {title} {\bibinfo {title}
  {{Dancing in the dark: galactic properties trace spin swings along the cosmic
  web}},\ }\href {https://doi.org/10.1093/mnras/stu1227} {\bibfield  {journal}
  {\bibinfo  {journal} {\mnras}\ }\textbf {\bibinfo {volume} {444}},\ \bibinfo
  {pages} {1453} (\bibinfo {year} {2014})},\ \Eprint
  {https://arxiv.org/abs/1402.1165} {arXiv:1402.1165 [astro-ph.CO]}
  \BibitemShut {NoStop}%
\bibitem [{\citenamefont {{Foreman-Mackey}}\ \emph {et~al.}(2013)\citenamefont
  {{Foreman-Mackey}}, \citenamefont {{Hogg}}, \citenamefont {{Lang}},\ and\
  \citenamefont {{Goodman}}}]{Foreman-Mackey2013}%
  \BibitemOpen
  \bibfield  {author} {\bibinfo {author} {\bibfnamefont {D.}~\bibnamefont
  {{Foreman-Mackey}}}, \bibinfo {author} {\bibfnamefont {D.~W.}\ \bibnamefont
  {{Hogg}}}, \bibinfo {author} {\bibfnamefont {D.}~\bibnamefont {{Lang}}}, and\
  \bibinfo {author} {\bibfnamefont {J.}~\bibnamefont {{Goodman}}},\ }\bibfield
  {title} {\bibinfo {title} {{emcee: The MCMC Hammer}},\ }\href
  {https://doi.org/10.1086/670067} {\bibfield  {journal} {\bibinfo  {journal}
  {\pasp}\ }\textbf {\bibinfo {volume} {125}},\ \bibinfo {pages} {306}
  (\bibinfo {year} {2013})},\ \Eprint {https://arxiv.org/abs/1202.3665}
  {arXiv:1202.3665 [astro-ph.IM]} \BibitemShut {NoStop}%
\bibitem [{\citenamefont {Binney}(1992)}]{Binney1992}%
  \BibitemOpen
  \bibfield  {author} {\bibinfo {author} {\bibfnamefont {J.}~\bibnamefont
  {Binney}},\ }\bibfield  {title} {\bibinfo {title} {{Warps}},\ }\href
  {https://doi.org/10.1146/annurev.aa.30.090192.000411} {\bibfield  {journal}
  {\bibinfo  {journal} {Annu. Rev. Astron. Astrophys.}\ }\textbf {\bibinfo
  {volume} {30}},\ \bibinfo {pages} {51} (\bibinfo {year} {1992})}\BibitemShut
  {NoStop}%
\bibitem [{\citenamefont {{LSST Science Collaboration}}\ \emph
  {et~al.}(2009)\citenamefont {{LSST Science Collaboration}}, \citenamefont
  {{Abell}}, \citenamefont {{Allison}}, \citenamefont {{Anderson}} \emph
  {et~al.}}]{LSST2009}%
  \BibitemOpen
  \bibfield  {author} {\bibinfo {author} {\bibnamefont {{LSST Science
  Collaboration}}}, \bibinfo {author} {\bibfnamefont {P.~A.}\ \bibnamefont
  {{Abell}}}, \bibinfo {author} {\bibfnamefont {J.}~\bibnamefont {{Allison}}},
  \bibinfo {author} {\bibfnamefont {S.~F.}\ \bibnamefont {{Anderson}}},  \emph
  {et~al.},\ }\bibfield  {title} {\bibinfo {title} {{LSST Science Book, Version
  2.0}},\ }\href@noop {} {\bibfield  {journal} {\bibinfo  {journal} {arXiv
  e-prints}\ ,\ \bibinfo {eid} {arXiv:0912.0201}} (\bibinfo {year} {2009})},\
  \Eprint {https://arxiv.org/abs/0912.0201} {arXiv:0912.0201 [astro-ph.IM]}
  \BibitemShut {NoStop}%
\bibitem [{\citenamefont {{Spergel}}\ \emph {et~al.}(2013)\citenamefont
  {{Spergel}}, \citenamefont {{Gehrels}}, \citenamefont {{Breckinridge}},
  \citenamefont {{Donahue}} \emph {et~al.}}]{Spergel2013}%
  \BibitemOpen
  \bibfield  {author} {\bibinfo {author} {\bibfnamefont {D.}~\bibnamefont
  {{Spergel}}}, \bibinfo {author} {\bibfnamefont {N.}~\bibnamefont
  {{Gehrels}}}, \bibinfo {author} {\bibfnamefont {J.}~\bibnamefont
  {{Breckinridge}}}, \bibinfo {author} {\bibfnamefont {M.}~\bibnamefont
  {{Donahue}}},  \emph {et~al.},\ }\bibfield  {title} {\bibinfo {title}
  {{WFIRST-2.4: What Every Astronomer Should Know}},\ }\href@noop {} {\bibfield
   {journal} {\bibinfo  {journal} {arXiv e-prints}\ ,\ \bibinfo {eid}
  {arXiv:1305.5425}} (\bibinfo {year} {2013})},\ \Eprint
  {https://arxiv.org/abs/1305.5425} {arXiv:1305.5425 [astro-ph.IM]}
  \BibitemShut {NoStop}%
\bibitem [{\citenamefont {{Laureijs}}\ \emph {et~al.}(2012)\citenamefont
  {{Laureijs}}, \citenamefont {{Gondoin}}, \citenamefont {{Duvet}},
  \citenamefont {{Saavedra Criado}}, \citenamefont {{Hoar}}, \citenamefont
  {{Amiaux}}, \citenamefont {{Augu{\`e}res}}, \citenamefont {{Cole}},
  \citenamefont {{Cropper}}, \citenamefont {{Ealet}}, \citenamefont
  {{Ferruit}}, \citenamefont {{Escudero Sanz}}, \citenamefont {{Jahnke}},
  \citenamefont {{Kohley}}, \citenamefont {{Maciaszek}}, \citenamefont
  {{Mellier}}, \citenamefont {{Oosterbroek}}, \citenamefont {{Pasian}},
  \citenamefont {{Sauvage}}, \citenamefont {{Scaramella}}, \citenamefont
  {{Sirianni}},\ and\ \citenamefont {{Valenziano}}}]{Laureijs:2012}%
  \BibitemOpen
  \bibfield  {author} {\bibinfo {author} {\bibfnamefont {R.}~\bibnamefont
  {{Laureijs}}}, \bibinfo {author} {\bibfnamefont {P.}~\bibnamefont
  {{Gondoin}}}, \bibinfo {author} {\bibfnamefont {L.}~\bibnamefont {{Duvet}}},
  \bibinfo {author} {\bibfnamefont {G.}~\bibnamefont {{Saavedra Criado}}},
  \bibinfo {author} {\bibfnamefont {J.}~\bibnamefont {{Hoar}}}, \bibinfo
  {author} {\bibfnamefont {J.}~\bibnamefont {{Amiaux}}}, \bibinfo {author}
  {\bibfnamefont {J.-L.}\ \bibnamefont {{Augu{\`e}res}}}, \bibinfo {author}
  {\bibfnamefont {R.}~\bibnamefont {{Cole}}}, \bibinfo {author} {\bibfnamefont
  {M.}~\bibnamefont {{Cropper}}}, \bibinfo {author} {\bibfnamefont
  {A.}~\bibnamefont {{Ealet}}}, \bibinfo {author} {\bibfnamefont
  {P.}~\bibnamefont {{Ferruit}}}, \bibinfo {author} {\bibfnamefont
  {I.}~\bibnamefont {{Escudero Sanz}}}, \bibinfo {author} {\bibfnamefont
  {K.}~\bibnamefont {{Jahnke}}}, \bibinfo {author} {\bibfnamefont
  {R.}~\bibnamefont {{Kohley}}}, \bibinfo {author} {\bibfnamefont
  {T.}~\bibnamefont {{Maciaszek}}}, \bibinfo {author} {\bibfnamefont
  {Y.}~\bibnamefont {{Mellier}}}, \bibinfo {author} {\bibfnamefont
  {T.}~\bibnamefont {{Oosterbroek}}}, \bibinfo {author} {\bibfnamefont
  {F.}~\bibnamefont {{Pasian}}}, \bibinfo {author} {\bibfnamefont
  {M.}~\bibnamefont {{Sauvage}}}, \bibinfo {author} {\bibfnamefont
  {R.}~\bibnamefont {{Scaramella}}}, \bibinfo {author} {\bibfnamefont
  {M.}~\bibnamefont {{Sirianni}}}, and\ \bibinfo {author} {\bibfnamefont
  {L.}~\bibnamefont {{Valenziano}}},\ }\bibfield  {title} {\bibinfo {title}
  {{Euclid: ESA's mission to map the geometry of the dark universe}},\ }in\
  \href {https://doi.org/10.1117/12.926496} {\emph {\bibinfo {booktitle} {Space
  Telescopes and Instrumentation 2012: Optical, Infrared, and Millimeter
  Wave}}},\ \bibinfo {series} {\procspie}, Vol.\ \bibinfo {volume} {8442}\
  (\bibinfo {year} {2012})\ p.\ \bibinfo {pages} {84420T}\BibitemShut {NoStop}%
\bibitem [{\citenamefont {{Cyr-Racine}}\ and\ \citenamefont
  {{Sigurdson}}(2013)}]{Cyr-Racine2013}%
  \BibitemOpen
  \bibfield  {author} {\bibinfo {author} {\bibfnamefont {F.-Y.}\ \bibnamefont
  {{Cyr-Racine}}}and\ \bibinfo {author} {\bibfnamefont {K.}~\bibnamefont
  {{Sigurdson}}},\ }\bibfield  {title} {\bibinfo {title} {{Cosmology of atomic
  dark matter}},\ }\href {https://doi.org/10.1103/PhysRevD.87.103515}
  {\bibfield  {journal} {\bibinfo  {journal} {\prd}\ }\textbf {\bibinfo
  {volume} {87}},\ \bibinfo {eid} {103515} (\bibinfo {year} {2013})},\ \Eprint
  {https://arxiv.org/abs/1209.5752} {arXiv:1209.5752 [astro-ph.CO]}
  \BibitemShut {NoStop}%
\end{thebibliography}%

\appendix
 
\section*{Appendix: Results in terms of the momentum transfer cross section}\label{sec:app}

Here we show the results of our work in terms of the momentum transfer cross section, to facilitate comparison with the literature. We begin by showing the theoretical formalism when using this cross section, and then display the results. The methods are the same as in the main text. We use Planck units in this appendix, except where we believe there may be confusion.

\subsection{Theory}
The momentum transfer cross section describes the momentum transferred between particles during an interaction, which is the process directly responsible for generating the drag. It is given by \citep{Kahlhoefer2014}:
\begin{equation}\label{eqn-sigmaT}
    \sigma_T = 4\pi \int_0^1 \frac{d\sigma}{d\Omega} (1 -\cos\theta)~ d(\cos\theta)\; ,
\end{equation}
where $\theta$ is the center of mass scattering angle, as in the main text. 

We will now give our drag force equations for each of the interaction types in terms of $\sigma_T$. We begin with the long-range interaction cross section, without any regularization term \citep{Kummer2017,Tulin2018}:
\begin{equation}
    \frac{d\sigma}{d\Omega} = \frac{\alpha^2_{\rm{DM}}}{m^2_{\rm{DM}}v^4\sin^4\theta} \; ,
\end{equation}
where we explicitly define the dependence on the coupling constant $\alpha_{\rm{DM}}$ and the mass of the DM particle, $m_{\rm{DM}}$.
The momentum transfer cross section for this differential cross section diverges at small scattering angles. In the main text, we used an extra $\sin\theta$ in the numerator to counteract this issue. Here, we will follow the normal procedure of introducing a Debye cut-off \citep[e.g,][]{Kahlhoefer2014, Cyr-Racine2013}. This sets a minimum scattering angle of:
\begin{equation}
    \theta_{\rm{min}} = \frac{4 \alpha_{\rm{DM}}}{\lambda_{\rm{De}} m_{\rm{DM}} v^2} \; ,
\end{equation}
where $v$ is the relative velocity between the halo and the background DM particles and $\lambda_{\rm{De}}$ is the Debye length. This is related to the total dark matter density near the particles, $\rho$, and other parameters by: $\lambda_{\rm{De}} = m_{\rm{DM}}v/(4\sqrt{\pi \alpha_{\rm{DM}} \rho})$. Using Equation~\ref{eqn-generaldrag}, we find that the acceleration due to the drag force should be:
\begin{equation}
    \vec{a}_{\rm{drag}} = -\frac{\sigma_T(v)}{2m_{\rm{DM}}} \rho_{\rm{bg}} v^2\ \hat{v} \; ,
\end{equation}
where we indicate explicitly the velocity-dependence of $\sigma_T$. Following Ref.~\cite{Kahlhoefer2014}, we can solve for the momentum transfer cross section as a function of $\theta_{\rm{min}}$:
\begin{equation}
    \sigma_T(v) = \frac{16\pi \alpha_{\rm{DM}}^2}{m^2_{\rm{DM}}} \frac{1}{v^4}  \left[ 1 - 2\log (\theta_{\rm{min}}/2) \right] \; .
\end{equation}

Substituting in the definition of $\theta_{\rm{min}}$ gives the explicit equation for $\sigma_T$ of:
\begin{equation}
    \sigma_T(v) = \frac{16\pi \alpha_{\rm{DM}}^2}{m^2_{\rm{DM}}} \frac{1}{v^4}  \left[ 1 - 2\log \left( \frac{8 \sqrt{\pi \alpha^3 \rho}}{m^2_{\rm{DM}} v^3}\right) \right] \; .
\end{equation}
We define $\log \Lambda (v) \equiv \log \left( \frac{8\sqrt{\pi \alpha_{\rm{DM}}^3 \rho}}{m^2_\text{DM} v^3}\right)$,\footnote{Including physical constants the equation is $\Lambda (v) = \left(8\sqrt{\pi \hbar^3c^5 \alpha_{\rm{DM}}^3 \rho}\right)/\left(m^2_\text{DM} (v/c)^3\right)$. This is the analog of the Coulomb logarithm in Rutherford scattering.} which is typically $\mathcal{O}(10^2)$ \citep{Kahlhoefer2014}. We now re-write the cross section in terms of $\sigma_T$ at a reference velocity $v_0$, which we take to be $300$ km/s:
\begin{equation}
    \sigma_T(v) = \sigma_T(v_0) \left(\frac{v_0}{v}\right)^4 \frac{1-2\log \Lambda(v)}{1-2\log\Lambda(v_0)}.
\end{equation}

We would like to continue without setting a dark matter particle mass or coupling constant. Thus, we want to approximate the logarithm in such a way that is appropriate for many possible values of these constants. First, let us consider the `typical' values for these constants. Ref.~\cite{Kahlhoefer2014} use $\alpha_{\rm{DM}} = 10^{-2}$ and $m_{\rm{DM}} = 1~\rm{TeV}$. We take the average dark matter density to be the average halo density at the scale radius of our halos: $\rho = 1.0\e{8}~M_{\odot}/{\rm{kpc}}^3$. Then our average logarithm value is: $\Lambda(v) = 4.4\e{-29}/(v/c)^3$. For $v = 300~\rm{km/s}$, the full logarithmic factor is: $1-2\log \Lambda(v) \sim 90$. We check this for a wide range of possible $\alpha_{\rm{DM}}$, $\rho$, and $m_{\rm{DM}}$, and we find that this expression is almost always $50 - 120$ (where most of this range come from the DM particle mass from $1~\rm{MeV} - 100~\rm{TeV}$). We also check that for $v = 10 - 5000~\rm{km/s}$, this factor ranges from $1-2\log \Lambda(v) = 70 - 120$. Finally, the ratio of the logarithmic factors is almost entirely independent of the parameters $\alpha_{\rm{DM}}$, $\rho$, $m_{\rm{DM}}$. For the full parameter ranges considered above and $v = 2v_0$, we find the ratio to be: $1.05-1.1$. Thus, the ratio of logarithmic factors is $\sim 1$ and it is mainly dependent on the ratio of the velocities. We will set the constants to the typical values above and just consider the effect of the velocities on the logarithmic factor. In other words, we set:  $\Lambda(v) = 4.4\e{-29}/(v/c)^3$. We find that this gives an adequate representation of the full range of possible values.

Our drag equation for long-range interactions is therefore:
\begin{equation}
    \vec{a}_{\rm{drag}} = -\frac{\sigma_{T}(v_0)}{2m_{\rm{DM}}} \rho_{\rm{bg}} v^2 \left(\frac{v_0}{v}\right)^4\frac{1-2\log \Lambda(v)}{1-2\log\Lambda(v_0)}\ \hat{v} \; ,
\end{equation}


For the contact interaction case, the relation between total and momentum transfer cross sections is simply $\tilde{\sigma} = 2\sigma_T$, where neither cross section has a velocity dependence. Our drag equation in this case is just:
\begin{equation}
    \vec{a}_{\rm{drag}} = -\frac{\sigma_{T}}{2m_{\rm{DM}}} \rho_{\rm{bg}} v^2\ \hat{v} \; .
\end{equation}

Finally, in the intermediate case, we again want to just interpolate between the two cases. We use:

\begin{equation}\label{eqn:app_m}
    \vec{a}_{\rm{drag}} = -\frac{\sigma_{T}(v_0)}{2m_{\rm{DM}}} \rho_{\rm{bg}} v^2 \left(\frac{v_0}{v}\right)^m\ \hat{v} \; .
\end{equation}
We do not include the logarithmic term from the Rutherford scattering formula because it is not a generic feature of finite mediator mass models.

Note that our expressions for the evaporation and velocity dispersion terms do not change and we include these effects in what follows.

\begin{table*}[!htb]
\begin{tabular}{|l|l|l||c|c|}
\hline
\hline
 Assumed Velocity & Evaporation? & Dispersion? & 68\% Upper Limit & 95\% Upper Limit\\ & & & $\rm{cm}^2/\rm{g}$ & $\rm{cm}^2/\rm{g}$ \\
 \hline
  \multirow{2}{*}{$v = 300 \: \rm{km/s}$} & N/A & - & $0.10$ & 0.21\\
  {} & N/A & \checkmark & $0.13$ & $0.29$\\
 \cline{1-5}
 \multirow{2}{*}{$v = v_{\rm{CF3}}$} & N/A & - & $0.03$& $0.07$\\
 {} & N/A & \checkmark & $0.21$ & $0.51$\\
 \hline
 \hline
\end{tabular}
\caption{Limits on the self-interaction momentum transfer cross section $\sigma_T(300 \: \text{km/s})/m_\text{DM}$ for long-range interactions, for different assumed galaxy velocities}
\label{tab:app_results}
\end{table*}

\begin{figure}
    \centering
    \includegraphics[width=0.5\textwidth]{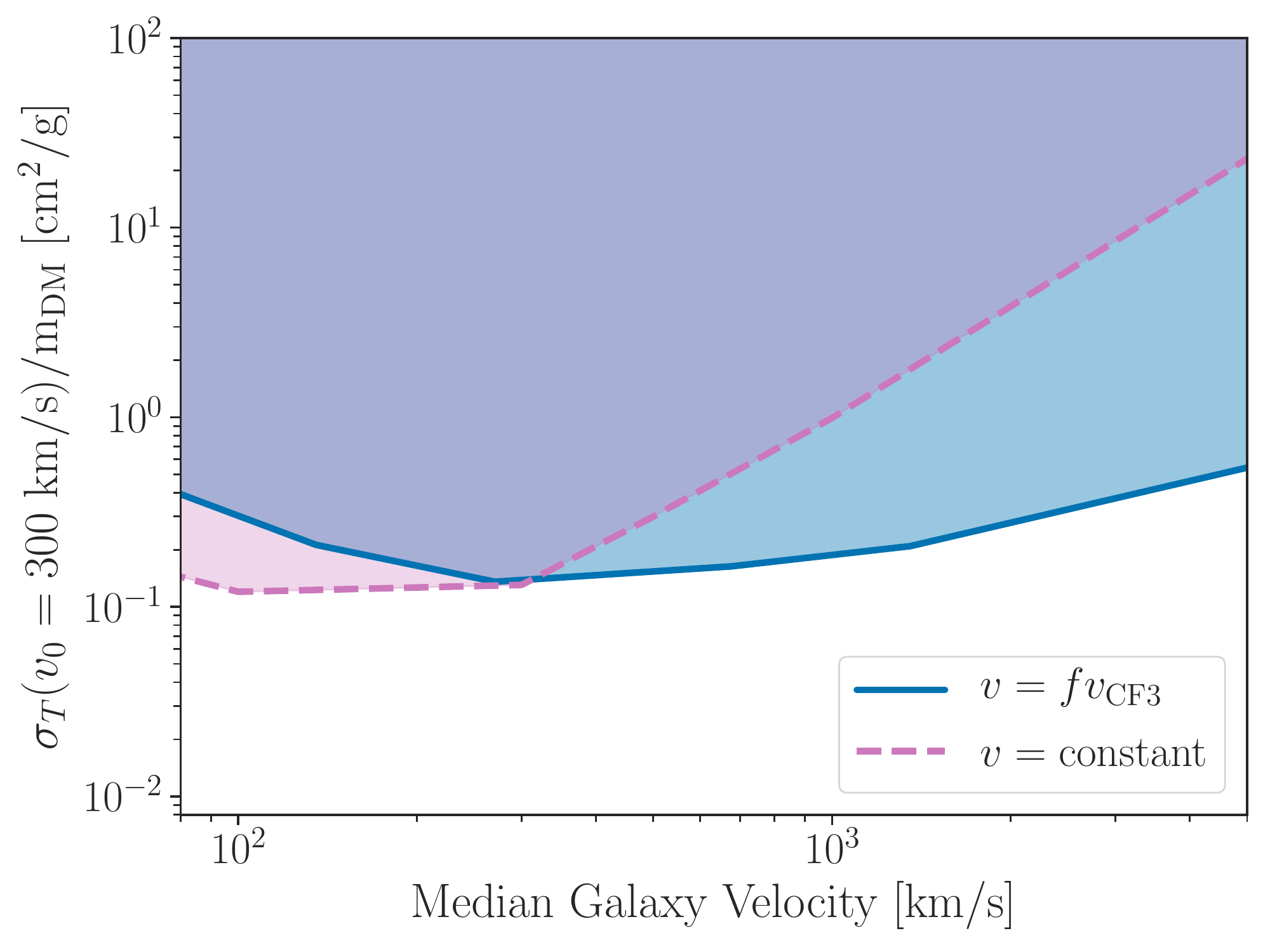}
    \caption{68\% upper limits on the momentum transfer cross section at 300 km/s, assuming a long-range interaction and plotted against the median assumed galaxy velocity. We show limits assuming either that all galaxies have the same relative velocities (dashed pink), or that they have velocities proportional to their CF3 velocities (solid blue).}
    \label{fig:app_long}
\end{figure}

\subsection{Results}
Our results for the long-range case are given in Table~\ref{tab:app_results} and Figure~\ref{fig:app_long}. Note that there is an $\mathcal{O}(10^{12})$ difference between these limits and the ones given in the main text. This occurs because the main difference in our equations for the drag in the main text and in this appendix is a factor $(c/v_0)^4 \sim 10^{12}$. The logarithmic term also depends on the velocity and changes the shape of the constraints slightly.

\begin{figure}[!htb]
    \centering
    \includegraphics[width=0.5\textwidth]{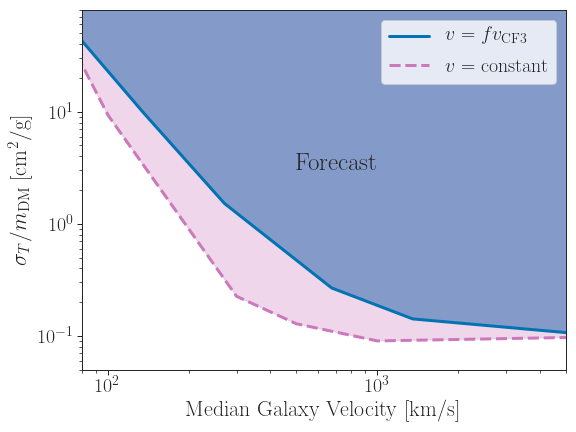}
    \caption{Forecasted 68\% upper limits on the momentum transfer cross section assuming a contact interaction versus the median assumed velocity, assuming a sample similar to ours but in environments where multi-streaming and the fluid approximation obtain. We show limits assuming either that all galaxies have the same relative velocities (pink), or that they have velocities proportional to their CF3 velocities (blue).}
    \label{fig:app_contact}
\end{figure}

Our forecasted contact interaction  limits are given in Figure~\ref{fig:app_contact}. These are twice as tight as the main text constraints since $\tilde{\sigma} = 2\sigma_T$ for the contact case.
\vspace{0.1cm}

\begin{figure}[!htb]
    \centering
    \includegraphics[width=0.5\textwidth]{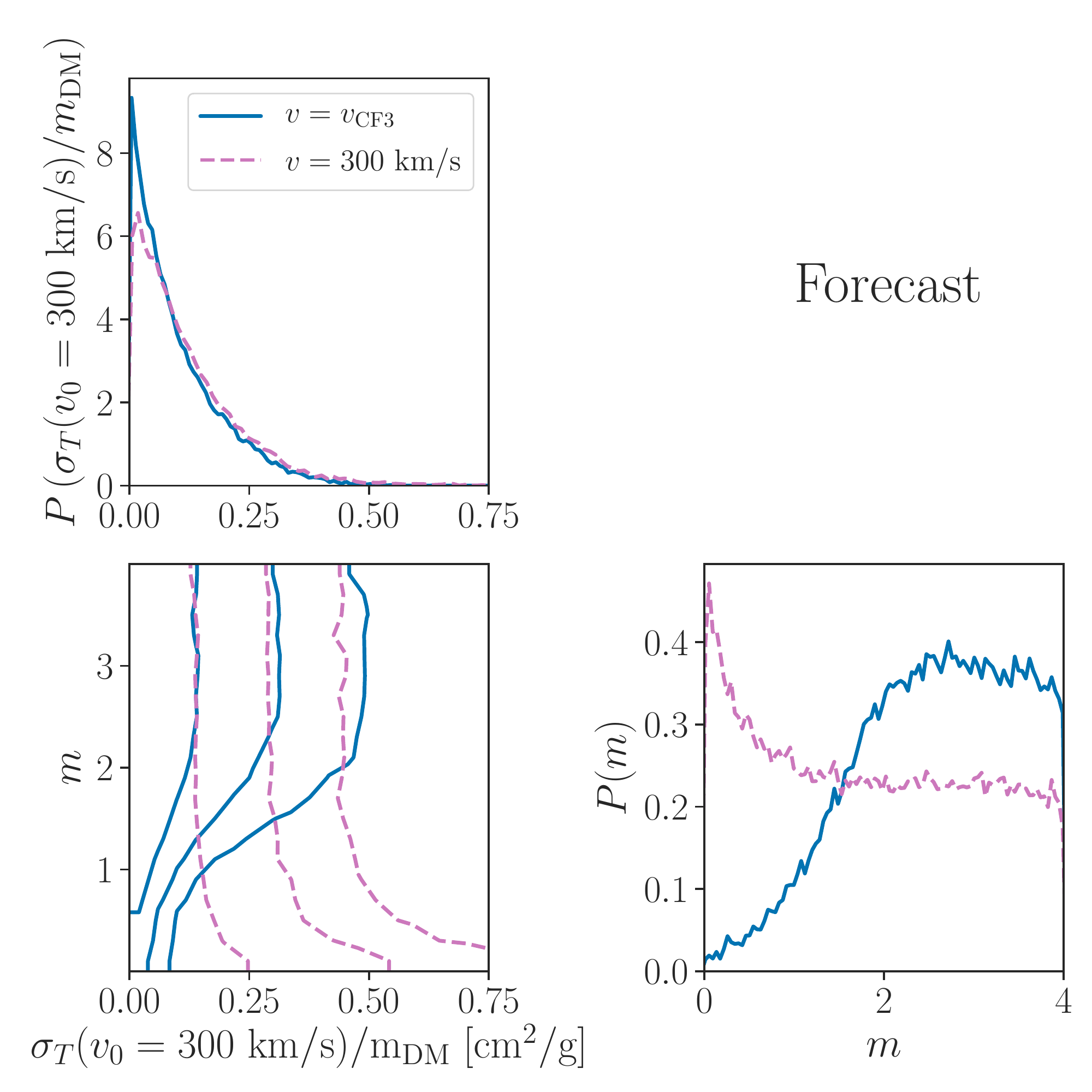}
    \caption{Forecasted corner plot for an intermediate-range DM self-interaction, assuming a sample similar to ours but in environments where multi-streaming and the fluid approximation obtain. We show our limits assuming all galaxies have $v=300~\rm{km/s}$ (pink) and assuming they have velocities set by their CF3 velocities (blue). $m$ determines the dependence of $a_\text{drag}$ on the relative velocity of the halo and background (Eq.~\ref{eqn:app_m}).}
    \label{fig:app_interp}
\end{figure}

Finally, our forecasted intermediate-range interactions are given in Figure~\ref{fig:app_interp}. Unlike in the main text, the cross section does not vary as widely and we are able to use a flat linear prior for the cross section. Thus, the contours are the true $1\sigma$, $2\sigma$, and $3\sigma$ upper bounds. Note that in the case of constant $v = 300$ km/s, sensitivity to $m$ arises only due to the effect of evaporation (Eq.~\ref{eq:evap}); without this the $m$ posterior would be flat. The mean CF3 velocity significantly exceeds $v_0$, so larger $m$ reduces $a_\text{drag}$ and hence allows larger values of $\sigma_T$. This access to a greater range of the $\sigma_T$ prior is responsible for the rise in the marginalised $m$ posterior towards larger values.

\end{document}